\documentclass[a4paper,12pt]{article}
\pdfoutput=1 %

\usepackage{jheppub} %

\usepackage{graphicx}%
\usepackage{bm}%
\usepackage{hyperref}%
\usepackage{amsmath}
\usepackage{amssymb}
\usepackage{graphicx,amssymb,amsmath,color,bm}
\usepackage{hyperref}
\usepackage{subcaption}
\usepackage{float}

\newcommand{\beq}{\begin{equation}}
\newcommand{\eeq}{\end{equation}}
\def\bea{\begin{eqnarray}}
\def\eea{\end{eqnarray}}
\def\beal{\begin{align}}
\def\eal{\end{align}}
\newcommand{\bei}{\begin{itemize}}
\newcommand{\eei}{\end{itemize}}
\newcommand{\bmat}{\begin{matrix}}
\newcommand{\emat}{\end{matrix}}

\newcommand{\Fig}[1]{Fig.~\ref{#1}}
\newcommand{\Eq}[1]{Eq.~(\ref{#1})}
\newcommand{\Sec}[1]{Sec.~\ref{#1}}
\newcommand{\App}[1]{Appendix~\ref{#1}}

\def\={\,=\,}
\def\+{\,+\,}
\def\-{\,-\,}

\def\Gpc{\text{Gpc}}
\def\Mpc{\text{Mpc}}

\def\kp{k_\parallel}
\def\kv{k_\perp}
\def\bk{\bm{k}}

\def\SNR{{\rm SNR}}
\def\Var{{\rm Var}}

\title{Probing small-scale power spectrum with gravitational-wave diffractive lensing}

\author[a]{Sungjung Kim,}
\affiliation[a]{Center for Theoretical Physics, Department of Physics and Astronomy, \\Seoul National University, Seoul 08826, Korea}
\emailAdd{sjkimphya@gmail.com}

\author[b]{Han Gil Choi,}

\affiliation[b]{Cosmology, Gravity and Astroparticle Physics Group, Center for Theoretical Physics of the Universe,
Institute for Basic Science (IBS), Daejeon 34126, Korea}
\emailAdd{hgchoi1w@gmail.com}

\author[a,c]{Sunghoon Jung}
\affiliation[c]{Astronomy Research Center, Department of Physics and Astronomy,\\ 
Seoul National University, Seoul 08826, Korea}
\emailAdd{sunghoonj@snu.ac.kr}

\abstract{
We develop a novel way to probe subgalactic-scale matter distribution with diffractive lensing on gravitational waves. Five-year observations from Einstein Telescope and DECIGO are expected to probe $k= 10^5\sim 10^8 \,{\rm Mpc}^{-1}$ down to $P(k) = 10^{-16} \sim 10^{-14} \,{\rm Mpc}^3$ level. These results can be interpreted in terms of primordial black holes in the range $M_{\rm PBH} \gtrsim 10^{-3}M_\odot$ down to $f_{\rm PBH} = 10^{-6}$ level, or QCD axion minihalos in the range $m_a = 10^{-3} \sim 10^{-12} \,{\rm eV}$. 
A key result of the paper is the approximate relation between the scale $k$ and the gravitational wave frequency $f$, derived in an ensemble of `multi-lensing' events. This relation enables direct measurement of the power spectrum at specific scales, with sensitivities characterized by model-independent kernels $\delta P(k)$. Additionally, we delineate the statistical properties of `multi-lensing' based on the `Fresnel number' $N_F$. When $N_F \gtrsim {\cal O}(1)$, the statistical significance can be approximately calculated by Variance of lensing effects, which is directly related to the power spectrum among other moments of matter distribution. 
}

\begin{document} 
\maketitle

\newpage
\section{Introduction}

Dark matter remains one of the biggest mysteries in the universe. Although its evidences are clear from cosmic microwave background (CMB) and galaxy rotation curves among many others, various searches of dark matter via direct, indirect, collider, and astrophysical probes all have failed to discover them. One of the regimes that have not been probed well is subgalactic-scale structures of dark matter. There are many possibilities that allow abundant such structures, such as primordial black holes or axion minihalos among many others. There are also many scenarios that suppress small-scale structures, such as long freestreaming length and self-interaction of dark matter. The search of subgalactic-scale structures is challenging basically because their gravitational effects are tiny and they are not bright. 

Only recently, various theoretical proposals have been put forward to probe this subgalactic regime, e.g. using dark structures' gravitational perturbations on star kinematics~\cite{VanTilburg:2018ykj,Ando:2022tpj,Mondino:2023pnc,Graham:2023unf,Graham:2024hah}, and lensing on lights from pulsars~\cite{Bai:2018bej,Dror:2019twh,Lee:2020wfn}, supernova~\cite{Zumalacarregui:2017qqd}, bursts of radios and gamma-rays~\cite{Munoz:2016tmg,Katz:2018zrn,Jung:2019fcs,Xiao:2024qay,Gould:1992,Nemiroff:1995ak,Nemiroff:2001bp}, nearby stars~\cite{Niikura:2017zjd,Niikura:2019kqi,DeRocco:2023hij,EROS-2:2006ryy,CalchiNovati:2013jpj,Griest:2013aaa}, caustic crossing~\cite{Oguri:2017ock,Dai:2019lud}, and lensing on gravitational waves (GWs)~\cite{Jung:2017flg,Dai:2018enj,Oguri:2020ldf,GilChoi:2023ahp,Zumalacarregui:2024ocb,LIGOScientific:2021izm,LIGOScientific:2023bwz,Nakamura:1997sw}. They are expected to be sensitive to a wide range of subgalactic scales from $k \sim 10 \,{\rm Mpc}^{-1}$ down to asteroid masses ($M \sim 10^{-16} M_\odot$). Spectral distortions of CMB can also probe up to $k \sim 10^{4} \,{\rm Mpc}^{-1}$~\cite{Chluba:2012we}. 

Many small-scale probes rely on detecting single event with tiny effects that dark matter structures may exert on precisely measured probes. But statistical variance as a signal of randomness of small-scale overdensities (such as location, mass and size) have also been proposed~\cite{Oguri:2020ldf,Xiao:2024qay,Zumalacarregui:2024ocb,Cyr-Racine:2018htu}. In most cases, however, the scale of structures has to be inferred from best-fit or model-dependent analysis, not directly measured.

In this paper, we advocate that GW diffractive lensing is a powerful probe of subgalactic-scale structures, which also provide with approximate measurement of scales. The main properties that allow these are: (1) chirping GWs from binary mergers have characteristic frequency spectrum, (2) diffractive lensing is frequency-dependent, (3) the observable lensing effect on the GW is not mere amplification but the frequency-dependence of amplification, and (4) lastly but very importantly, the observable with GW frequency $f$ is most sensitive to a particular scale given by the Fresnel scale (as a function of $f$)~\cite{Choi:2021bkx}. We build upon a pioneering work on this subject~\cite{Oguri:2020ldf} and our own development in \cite{Choi:2021bkx}.

To probe small-scale structures which typically exert only weak gravitational effects, we statistically combine an ensemble of sub-critical GW events of lensing (which by themselves cannot claim detection but are clean enough). The combination not only simply increases the total significance (enabling the search that was not possible with single strong event), but also makes it possible to semi-directly indicate the $k$-scale of the mass distribution corresponding to the frequency $f$. In other words, statistical properties of GW lensing spectrum allow the measurement of small-scale overdensity distributions in the corresponding $k$ scale. The approximate relation between $f$ and $k$ is one of the main results of this paper. Then it becomes clear that Einstein Telescope(ET) and DECIGO (probing $f = 10^{-1} \sim 10^3$ Hz) will be sensitive to $k = 10^{5} \sim 10^8\,{\rm Mpc}^{-1}$.

Another critical effect to be accounted for is so called `multi-lensing'. By multi-lensing, we mean to consider all lenses or general mass distributions along the propagation. This is in contrast to usual previous studies of GW lensing, in which only a single strong lens is considered.
The potential benefits of multi-lensing in the wave-optics regime for gravitational wave    observations have been recognized only recently~\cite{Urrutia:2024pos, Zumalacarregui:2024ocb}.
After all, we will quantify the improvement from this consideration and delineate genuine events of multiple lensing.

To sum, all three effects are critical to probe subgalactic structures: frequency dependences of lensing and detection, statistical combination of sub-critical events, and possible multi-lensing along the line of sight. We discuss the relevance of each physics in different regions of the dark matter parameter space.

This paper is organized as follow. \Sec{sec:multilensing} formulates multi-lensing with relevant scales and statistical properties. \Sec{sec:likelihood} introduces the significance measure and the Gaussianity of total significance. \Sec{sec:method} introduces Monte-Carlo simulation methods. We present sensitivities in \Sec{sec:result}, along with statistical properties of multi-lensing. We summarize in \Sec{sec:summary}.

\section{Multi-lensing} \label{sec:multilensing}

{\bf Multi-lensing and Fresnel scales.}
The complex lensing amplification $F(f)$ defined as the ratio of lensed $h_L(f)$ and unlensed $h_0(f)$ waveforms
\beq
h_L(f) \= F(f) h_0(f)
\eeq
is given by a 3d path integral, treating each GW polarization as an independent scalar wave degree of freedom with Born approximation for weak gravitational potential~\cite{Takahashi:2005sxa,Takahashi:2005ug,Oguri:2020ldf},
\begin{align}
	F(f; \chi_s) -1 &\,\equiv\, \eta(f; \chi_s) \label{eq:F}\\
    &\,\simeq\, -{4\pi i f}
		\int \frac{d^3\bm{k}}{(2\pi)^3} \int d\chi_l \tilde{\Phi}(\bm{k}) e^{ik_\parallel \chi_l} \left(\exp\left[ -i\frac{\chi_l(\chi_s-\chi_l)}{4\pi f \chi_s}|\bm{k}_\perp|^2 \right] - 1\right) \nonumber\\
		&\,\equiv\, \int \! \frac{d^3\bm{k}}{(2\pi)^3} \int_0^{\chi_s} \! d\chi_l \; \tilde{\Phi}(\bm{k}) e^{i\kp\chi_l} g(f; \kv, \chi_l).
	\label{eq:etaPk}
\end{align}
$\eta$ is the (reduced) lensing amplification, $\tilde{\Phi}(\bm{k})$ the (lens) gravitational potential in the comoving Fourier space, and $\chi_{s,l}$ the comoving distances to the source and lenses.
\Eq{eq:etaPk} is our main technical equation that we use to calculate observables and sensitivities. 
This general expression incorporates both \emph{multi-lensing} along the line of sight and \emph{frequency dependence} of lensing (wave optics, or equivalently diffractive lensing). 
In this work, it is important to take into account both physics.

\begin{figure}[t]
	\centering
	\includegraphics[width=0.95\linewidth]{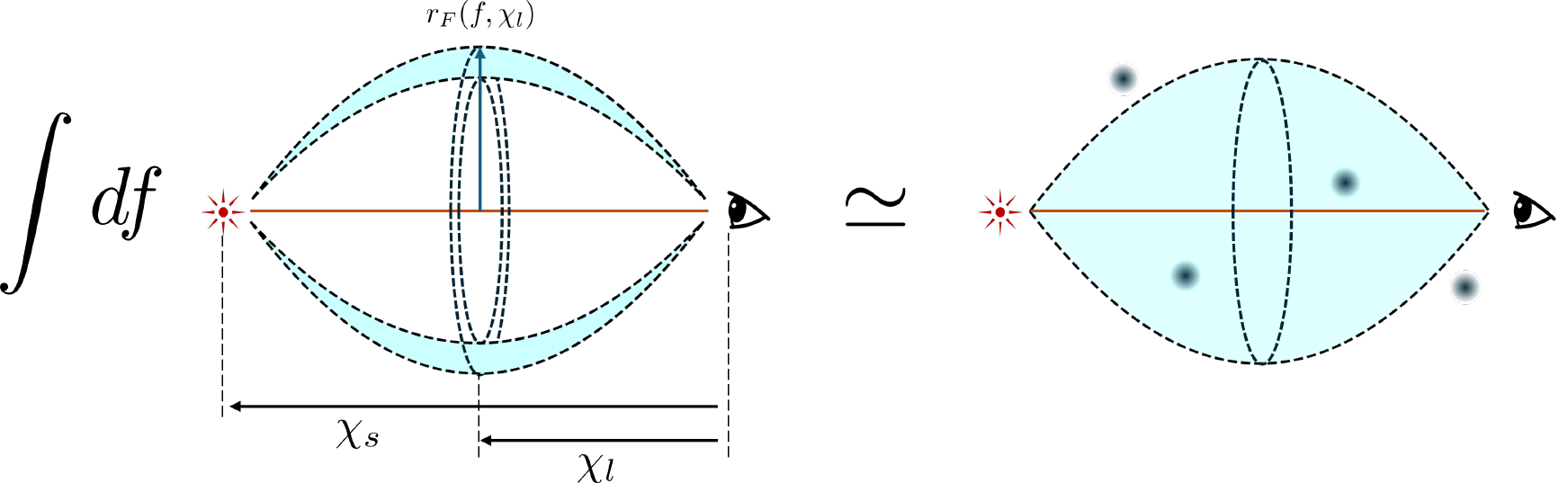}
	\caption{The \emph{observable} lensing effect on the GW with frequency $f$ is sensitive to the shear (the radial variation of mass density) in the Fresnel shell (shaded), defined with the Fresnel length $r_F(f,\chi_l)$ (\Eq{eq:kF}). The lensing significance of one event is its frequency integral, sweeping a range of shell and yielding the concept of the Fresnel volume and Fresnel number (\Eq{eq:fresnum}), which will be critical to statistical properties of multi-lensing and validity of the Central Limit Theorem for total likelihood; see \Sec{sec:mc} and \ref{sec:properties}. In this regime, the statistical Variance of GW observables is dominated by the power spectrum at a mid point $\chi_l \simeq \frac{\chi_s}{2}$ with $k=\bar{k}_F$; see \Eq{eq:var_approx}.}
	\label{fig:situation}
\end{figure}

Consider first diffractive lensing by a single lens -- the major subject of GW lensing so far. In this case, the 3-d integral is approximated by the 2d one on the lens plane. The lensing effect on the GW can then be described in terms of the enclosed mass density and its radial variation (called the shear $\gamma(r)$) around the radial distance of the Fresnel length scale $r=r_F(f)$. Notably, the shear plays a central role in GW diffractive lensing because the \emph{observable} lensing effect of GW with frequency $f$ is the frequency dependence of $F(f)$, namely $\eta'(f) = \frac{d\eta(f)}{d\ln f}$, not mere amplification, which arises from the average shear around $r_F(f)$~\cite{Choi:2021bkx,Jung:2022tzn}. Consequently, observing the lensing effect as a function of $f$ is equivalent to measuring the shear \emph{distribution}, thereby enabling both lensing detection and mass distribution measurements.

We can apply this intuition to the case of multi-lensing, or  lensing integrated along the line of sight calculated by \Eq{eq:etaPk}. As illustrated in \Fig{fig:situation}, the observable lensing effect on GW with $f$ is approximately sensitive to shears in the Fresnel shell region (defined by $r_F(f,\chi_l)$)
\beq
\eta'(f) \,\sim\, \int d\chi_l \,\gamma(r_F(f,\chi_l)).
\eeq
As with the shear, this measures the radial variation of density in the shell region as well as the average density enclosed by the shell.

However, unlike the single-lens case, the observable with $f$ is not sensitive to a single scale $r_F(f)$, but an integration of various scales $r_F(f,\chi_l)$ along the line of sight. Fortunately, we find that statistical properties of lensing effect, for example the variance Var($\eta'(f)$) in an ensemble of GW events is dominated by $\bar{r}_F(f) \equiv r_F(f, \chi_l=\chi_s/2)$ around a mid point. Precisely speaking, this is derived in the $k$-space; the Variance is dominated by the matter power at $k\simeq \bar{k}_F \equiv \pi/\bar{r}_F$
\bea
k_F(f,\chi_l) &\,\equiv\,& \frac{\pi}{r_F(f,\chi_l)} \= \sqrt{\frac{2\pi^3 f}{\chi_{\rm eff}}}, \label{eq:kF} \\
\bar{k}_F(f) &\,\equiv\,& k_F(f,\chi_l = \chi_s/2) \,\simeq\, 5.06 \times 10^6 \,{\rm Mpc}^{-1} \left( \frac{\chi_s}{\rm Gpc} \right)^{-1/2} \left( \frac{f}{\rm Hz} \right)^{1/2}.
\label{eq:bkF}\eea
Here, $\chi_{\rm eff} = \chi_l (\chi_s-\chi_l)/\chi_s$ is the effective lensing distance.

\medskip
{\bf Statistics of $\eta'$, dominated by $k\simeq \bar{k}_F$.}
The statistical properties of lensing effects are critical elements of this work. $\eta'$ varies event-by-event, and its statistical properties encode information of matter distributions at relevant scales. 

For example, the statistical Variances of $\eta$ and $\eta'$ (among other statistical moments) are analytically related to the power spectrum (more precisely $k^2 P(k)$ in $\ln k$-space) as follow
\begin{align}
	{\rm Var}(\eta)
	&\= \left< \eta(f) \eta(f)^* \right> \\
	&\= \int \frac{d^3\bm{k}}{(2\pi)^3} \frac{d^3\bm{k'}}{(2\pi)^3} \int_0^{\chi_s} d\chi_l  d\chi'_l
		\left< \tilde{\Phi}(\bm{k})^* \tilde{\Phi}(\bm{k'})\right>
		e^{i (\kp\chi_l-\kp'\chi'_l)} g(f;\kv, \chi_l)g^*(f;\kv', \chi'_l) \nonumber\\
	&\= \int \frac{d^3\bm{k}}{(2\pi)^3} \int d\chi_l d\chi'_l P_\Phi(k) e^{i\kp(\chi_l-\chi'_l)} g(f;\kv, \chi_l) g^*(f; \kv, \chi'_l)
    \nonumber\\
	&\,\simeq\, \frac{\chi_s^3}{60\pi}(4\pi G \bar{\rho})^2 \int d\ln \kv \, \cdot \frac{960\pi^2 f^2}{\kv^4\chi_s^2}
	\int_0^{\chi_s} \frac{d\chi_l}{\chi_s} \, a(\chi_l)^{-2} \kv^2  P(\kv) \left[ 1 - \cos\left( \frac{\chi_l(\chi_s-\chi_l)}{4\pi f \chi_s} \kv^2 \right)\right] \nonumber\\
	&\,\equiv\, \frac{\chi_s^3}{60\pi}(4\pi G \bar{\rho})^2 \int d\ln \kv\, \kv^2 P(\kv)_{z=0} \cdot \mathcal{G}_0(\ln \kv, \ln \bar{k}_F(f)),
\label{eq:var_eta} \end{align}
where $\left< \tilde{\Phi}^* \tilde{\Phi} \right> = (2\pi)^3 \delta(\bk-\bk') P_\Phi(k)$, $P_\Phi(k) = (4\pi G \bar{\rho}a^{-1}/k^2)^2 P(k)$, and the Limber's approximation $\int \frac{\kp}{2\pi}e^{i\kp(\chi-\chi')}f(k) \simeq \delta(\chi-\chi')f(\kv)$ are used. In the last line, all $\chi_l$-dependencies, including the scale factor $a(\chi)$ and  redshift $z$, are absorbed into ${\cal G}_0$, and after all ${\cal G}_0$ is dominated by $\chi_l \simeq \chi_s/2$ (hence, by $k\simeq \bar{k}_F$) as shown below and in \App{app:G1}; so the subscript $z=0$ appears only in this equation.
We will also drop the $\perp$ symbol and denote $k_\perp$ by $k$. See \cite{Oguri:2020ldf} for related calculations.

The Variance of \emph{observable} lensing effect, $\eta'$, is similarly given by (again $k_\perp \to k$)
\begin{align}
	{\rm Var} \left( \frac{d\eta(f)}{d \ln f} \right)
	&\,\simeq\, \frac{\chi_s^3}{2\pi} \int d\ln k \, k^2 P(k) \int_0^{\chi_s} d\chi_l \, \left( \frac{4\pi G\bar{\rho}}{a} \right)^2 \left|\frac{dg(f; k, \chi_l)}{d \ln f}\right|^2 \nonumber\\
	&\,\equiv\, \frac{\chi_s^3}{60\pi}(4\pi G \bar{\rho})^2 \int d\ln k \, k^2 P(k)_{z=0} \cdot \mathcal{G}_1(\ln k, \ln \bar{k}_F(f)).
	\label{eq:var_d_eta}
\end{align}
The kernel is dimensionless and normalized as $\int d \ln k \, \mathcal{G}_1(\ln k) \sim \mathcal{O}(1)$.
Both statistical Variances of $\eta$ and $\eta'$ are related to $\int \,d\ln k \, k^2 P(k)$, which can be understood as 2d polar integration of power.

We emphasize that the Variance may not directly yield detection sensitivities since higher moments of potentials and $\eta'$ distributions may all be needed. Throughout this paper, we discuss why Variances are particularly relevant and when they are most useful.

\begin{figure}[t]
	\centering
	\includegraphics[width=.49\linewidth]{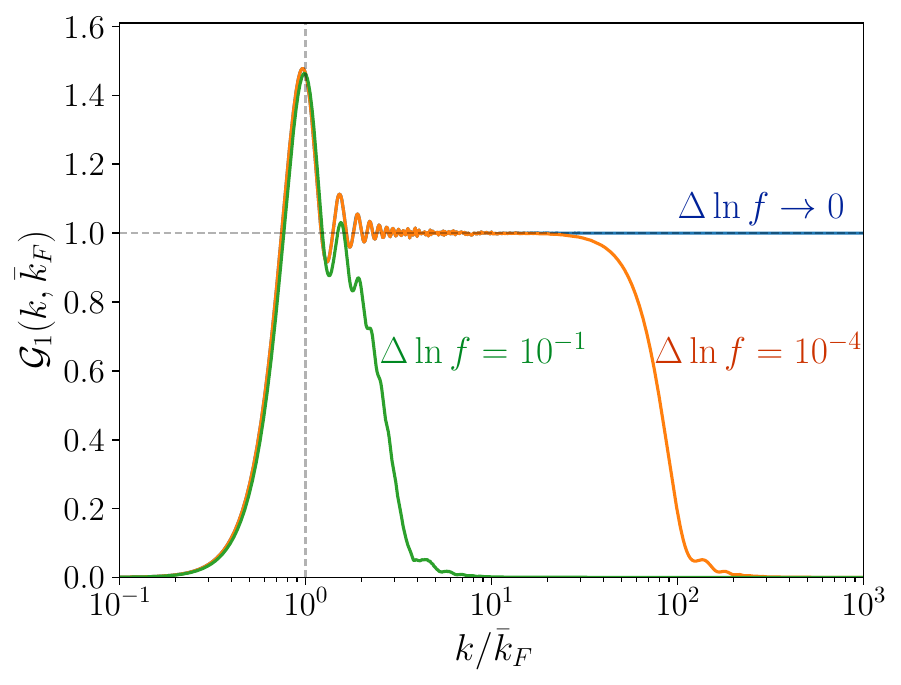}
    \includegraphics[width=0.49\linewidth]{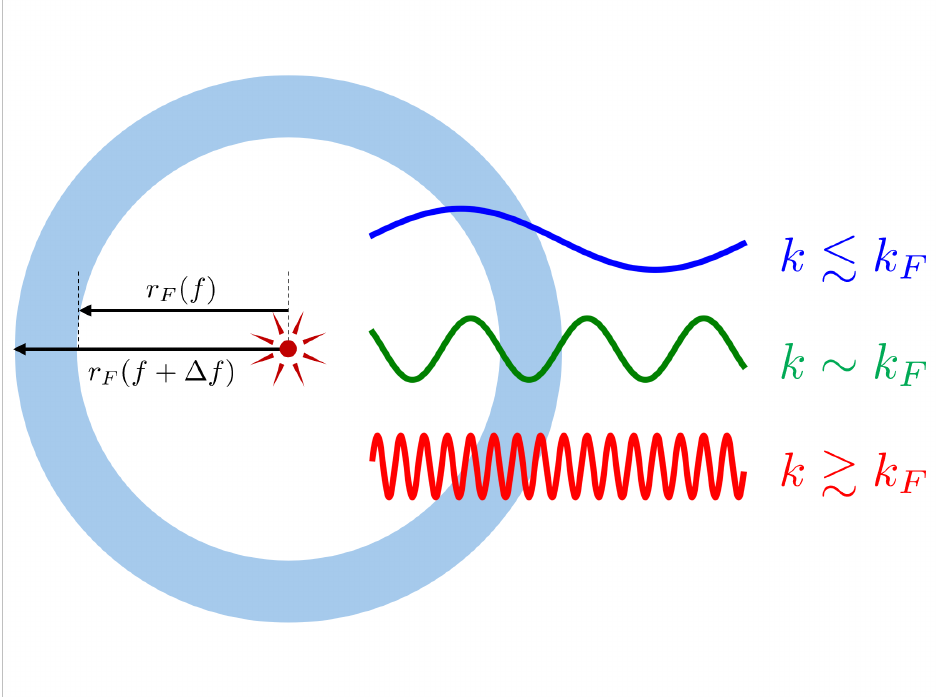}
	\caption{{\bf(Left):} $\mathcal{G}_1(k)$, the kernel for the GW observable Var($\eta'$) in \Eq{eq:var_d_eta}, receives contributions from all $k\gtrsim \bar{k}_F(f)$ for given $f$ (blue); the \emph{observable} is the enclosed mass density \emph{variation} within $r \simeq r_F(f)$. Practically, differentiated over a finite frequency width, the kernel is peaked at $k\sim \bar{k}_F$ (green), further illustrated in the {\bf right} panel. This implies that the intuition of single-lens diffraction in the $r$-space can be similarly applied to multi-lensing in the $k$-space.}
	\label{fig:G1}
\end{figure}

In any case, \Eq{eq:var_d_eta} provides powerful insight. As shown in \Fig{fig:G1} left, the kernel ${\cal G}_1(f,\ln k)$ is sharply peaked at $k\sim \bar{k}_F(f)$ (\Eq{eq:bkF}), which is the effective Fresnel scale corresponding to $f$ evaluated at $\chi_l = \chi_s/2$,
\beq
	{\cal G}_1(f,\ln k,\chi_l) \,\sim\, \delta (\ln k - \ln \bar{k}_F(f)).
\label{eq:G1delta}\eeq
Note that it is $\bar{k}_F(f)$, not $k_F(f,\chi_l)$, as an effective result of $\chi_l$ integral included in the definition of ${\cal G}_1$.
In general, no unique Fresnel scale exists in multi-lensing unlike the single-lens case, but this powerful property selects out a single scale $k \sim \bar{k}_F$. More discussions on \Fig{fig:G1} are given in the side remarks below.

Put in another way, this approximate property has two useful meanings: 
\bei
\item As in the single-lens case, the Fresnel scale $k_F(f,\chi_l)$ dominates lensing contribution to $f$ among all $k$, for the given $\chi_l$.
\item $\chi_l \simeq \chi_s/2$ (hence $k \simeq \bar{k}_F$) dominates the \emph{statistical} properties of lensing among all $\chi_l$ and $k_F(f,\chi_l)$. Thus, $k = \bar{k}_F(f) = k_F(f,\chi_l=\chi_s/2)$ is a useful parameter characterizing statistical properties of multi-lensing.
\eei

In all, the lensing effect on the GW with $f$ is dominated by the power at the corresponding $\bar{k}_F(f)$ as
\begin{align}
	{\rm Var} \left( \frac{d\eta(f)}{d \ln f} \right) 
	&\,\sim\, \mathcal{O}(1)\cdot \frac{\chi_s^3}{60\pi}(4\pi G \bar{\rho})^2 {\bar{k}_F^2 P(\bar{k}_F)}  \nonumber\\
	&\,\sim\, 2.97 \times 10^{-9} \left( \frac{\chi_s}{1{\rm Gpc}} \right)^3 \left( \frac{\bar{k}_F^2 P(\bar{k}_F)}{1{\rm Mpc}} \right).
	\label{eq:var_approx}
\end{align}
An ensemble of GW lensed events is most sensitive to the matter distribution at $\bar{k}_F$. 
In a sense, this is in accord with diffractive lensing by a single (diffuse) lens, in which the lensing effect with $f$ is determined dominantly by the mass density at distance $r \simeq r_F = k_F^{-1}$ from the lens center.

\medskip
{\bf Side remarks on ${\cal G}_0$ and ${\cal G}_1$.}
Eqs.~(\ref{eq:var_eta}) and (\ref{eq:var_d_eta}) are exact relations, not only for wave-optical diffraction but also for geometrical optics, and not only for multi-lensing but also for single-lensing. So we would like to make some general remarks, encompassing these regimes.

(1) In principle, ${\cal G}_1$ is a heaviside $\theta(k-\bar{k}_F)$ while ${\cal G}_0$ is an opposite heaviside $\theta(\bar{k}_F -k)$; see \Fig{fig:G1} and \Fig{fig:G0} in \App{app:G1}. ${\cal G}_1$ measures the variation of enclosed density, so only large $k$ modes contribute. On the other hand, ${\cal G}_0$ measures the total enclosed density (Gauss' theorem), so only small $k$ modes contribute (See the right panel in \Fig{fig:G1}). These two are consistent with each other, and these are realized by mathematical properties of $\int d\chi_l |g|^2$ and $\int d\chi_l |g'|^2$ in Eqs.~(\ref{eq:var_eta}) and (\ref{eq:var_d_eta}).

(2) However, practically, it might be more appropriate to differentiate ${\cal G}_1$ over a small finite frequency width $\Delta f$. The observable is then sensitive to the change of mass density by $r_F(f+\Delta f) \to r_F(f)$, hence to only those modes (green) that roughly fit the thickness of the annulus formed by the radii. Therefore, for $\Delta f/f \sim {\cal O}(1)$, the kernel ${\cal G}_1$ asymptotes to a peaked function, as shown in the left panel in Fig.~\ref{fig:G1}.

(3) The frequency $f$ appears only inside $g$ and $g'$ functions. This is one reason why only properties of ${\cal G}_{0,1}$ were relevant above. It also makes the overall $k^2 P(k)$ dependence (which appears outside ${\cal G}_{0,1}$) same in Var($\eta$) and Var($\eta'$) as well as same in wave-optics and geometrical-optics regimes. These were also obtained in \cite{Oguri:2020ldf} for Var($\eta'$) in diffractive lensing and in \cite{Xiao:2024qay} for Var($\eta$) in geometrical-optics lensing.

\section{Likelihood function, and its Gaussianity} \label{sec:likelihood}

\medskip
{\bf Gaussianity of $\ln \Lambda$ and Fresnel number.}
An important statistical property in this work is the probability distribution of the total log-likelihood of lensing
\beq
\ln \Lambda \= \sum_{i \,\in\,{\rm events}} \, \langle \ln \Lambda^i \rangle.
\eeq
It determines detection sensitivities; we judge that detection is possible if the probability for $\ln \Lambda$ to exceed the critical value $\ln \Lambda_c = 3$ is greater than 90\%. Thus, $\ln \Lambda$ probability distribution (not a single value) needs to be simulated. 

In particular, the (non-)Gaussianity of $\ln \Lambda$ distribution matters, for two reasons:
\bei
\item If Gaussian, the Variance of $\eta'$ in \Eq{eq:var_d_eta} can directly determine detection sensitivities (as will be discussed in \Eq{eq:lnlam_ave}). Otherwise, $\ln \Lambda$ distributions must be obtained by Monte-Carlo event simulations, as we will do. 
\item The Gaussianity is closely related to other critical properties of multi-lensing: the importance of sub-critical events (having $\ln \Lambda^i < \ln \Lambda_c$), and importance of multiple lenses along the line of sight. See \Sec{sec:properties}.
\eei

The total $\ln \Lambda$ distribution can be Gaussian even though individual $\ln \Lambda^i$ may not be. This is guaranteed by the Central Limit Theorem (CLT), if the number of individual contributions added is large enough; each lens and event are independent. The effective number of contributions for the CLT of $\ln \Lambda$ can be quantified by ``the Fresnel number'' $N_F$ (illustrated in \Fig{fig:situation}):
\beq
N_F \,\equiv\, \sum_{i \, \in \, {\rm events}} \,\textrm{Number of lenses within the Fresnel volume}.
\label{eq:fresnum} \eeq
This counts the number of lenses in the Fresnel volume (the 3d region bounded by the Fresnel shell), summed over whole events in the dataset. Thus, $N_F$ counts only those lenses that induce sizable diffractive lensing, in the whole event set. We will demonstrate in \Sec{sec:properties} that $N_F \gtrsim 1$ indeed is an important boundary for the validity of the CLT and Gaussianity of $\ln \Lambda$, as well as other properties of multi-lensing. Of course, $N_F$ and the Fresnel volume depend on the frequency, but this criteria approximately holds by using $f=f_0$, the frequency at which SNR contribution is maximal for the majority of events.

\medskip
{\bf Likelihood function.} The lensing likelihood of each event $i$ is expanded as
\beq
\langle \ln \Lambda^i \rangle \,\simeq\, \ln \Lambda_0^i + \ln \Lambda_2^i.
\label{eq:lnlami_exp} \eeq
The likelihood that we calculate is the one averaged over a noise ensemble (denoted by $\langle \cdot \rangle$), without simulating individual noise. The first term $\ln \Lambda_0^i$ is the widely used leading result (independent on individual noise), while the second term $\ln \Lambda_2^i$ includes leading degradation due to event-by-event noise variance. We then introduce event selections based on SNR$_i$ and $\ln \Lambda_2^i / \ln \Lambda_0^i$.

\medskip
{\bf Leading $\ln \Lambda_0^i$.}
Each event yields lensing likelihood $\ln \Lambda^i$ measured by the log of Bayes factor
\begin{align}
  \ln \Lambda^i \= -\frac{1}{2}(d-h_{L,BF}|d-h_{L,BF}) + \frac{1}{2}(d-h_{0,BF}|d-h_{0,BF}), 
  \label{eq:lnLambda}
\end{align}
where the best-fits by lensed and unlensed templates are compared. Data, $d(f) = h_L(f)+n(f)$, consists of lensed signal and random noise with spectral density $S_n(f)$. $(\cdot | \cdot)$ is the usual inner product $(a|b) \equiv 4 {\rm Re}\int d\ln f\, a^*(f) b(f) / S_n(f)$, and $\langle \cdot \rangle$ is the average under noise ensemble.

The Bayesian definition converges to the frequentist one, under noise ensemble average ($\langle {\cal O}(n) \rangle =0$, or $(n|h)/(h|h)\ll 1$)~\cite{Jung:2017flg,Dai:2018enj,GilChoi:2023ahp}
\bea
  \langle \ln \Lambda^i \rangle  \,\simeq\, \ln\Lambda_0^i &\,\equiv\,&  \langle \ln \Lambda^i_{{\cal O}(n^0)} \rangle \= \frac{1}{2} (h_L - h_{0,BF} | h_L - h_{0,BF} ),
  \label{eq:lnLambda0_def} \\
  \ln \Lambda_1^i &\,\equiv\,& \langle \ln \Lambda^i_{{\cal O}(n^1)} \rangle \= 0. 
\eea
This frequentist result can also be thought of as the chi-square measure of the unlensed best-fit. The convergence means that on average lensed templates fit the lensing signal well; we know that the signal is generated by lensing, not by other effects.

After the best-fit by unlensed waveforms $h_{\rm 0, template} = A_0 h_0 \,e^{i\phi_0} \,e^{i2\pi f t_c}$ with fitting parameters $\{ A_0, \phi_0, t_c \}$, this can be further approximated as~\cite{Choi:2021bkx}
\bea
\ln \Lambda^i_0  \,\simeq\, (\eta' h_0 | \eta' h_0 ) \,\sim\,  \SNR_i^2 \left| \frac{d\eta_i(f_0)}{d\ln f} \right|^2.
  \label{eq:lnLambda0_approx}
\eea
First of all, the fact that $\ln \Lambda^i$ is determined by $\eta'(f)$ significance shows that the observable effect is only frequency-dependent lensing effect; for example, the overall amplification will not be readily discerned from source distance variation. 

Secondly, this is the contribution of order $\sim {\cal O}(\SNR_i^2)$ arising at ${\cal O}(n^0)$. For the usual case of single event detection, this contribution is good enough since each signal is sufficiently strong. But in our case of multi-lensing, we also sum relatively weak signals, so subleading corrections shall be considered carefully.

\medskip
{\bf Subleading $\ln \Lambda_2^i$.} These are contributions that arise at $\langle {\cal O}(n^2) \rangle$, due to individual variation of noise.
There are two types of corrections. %
The first is the systematic bias of the best-fit; and the second is the individual error by noise variation added in quadrature over all events.  
The latter is statistical, whose mean is zero, hence can be improved with the amount of dataset as usual. %
So we do not include them in our work.  %
However, the former is the correction to $\langle \ln \Lambda\rangle$ so that this cannot be improved simply by increasing dataset.

The systematic bias turns out to arise from $t_c$ best-fit; while same effects from $A_0, \phi_0$ best-fits turn out to vanish.
For example from $A_0$ best-fit, $\langle {\cal O}(n^2) \rangle$ contributions read
\beq
\ln \Lambda_2 \,\equiv\, \langle \ln \Lambda_{{\cal O}(n^2)} \rangle  \,\ni\, \left\langle \frac{(n|h_L)^2}{(h_L|h_L)} -\frac{1}{2}\frac{(n|h_L)^2}{(h_L|h_L)} \,\-\, \frac{(n|h_0)^2}{(h_0|h_0)} +\frac{1}{2} \frac{(n|h_0)^2}{(h_0|h_0)} \right\rangle \= 0,
\label{eq:On2}\eeq
which vanishes due to $\langle (n|a)(b|n) \rangle = (b|a)$. Such corrections from $\phi_0$ best-fit also vanish in the same way. This is because the bias of the $A_0, \phi_0$ best-fits is same in lensed best-fit and unlensed best-fit, resulting in no change of lensing significance.  But such bias of $t_c$ best-fit do not cancel between lensed and unlensed best-fits; see \App{app:lnlam} for technical expressions.  We include this correction in our final results.

$\ln\Lambda_2$ is of order $\sim {\cal O}({\rm SNR}^0)$ (albeit some range of variation numerically) as can be deduced from each term in \Eq{eq:On2}, so this can even be larger than $\ln \Lambda_0$, which was $\sim {\cal O}({\rm SNR}^2)$. This will invalidate the expansion \Eq{eq:lnlami_exp}. We introduce event selections that discard such events.

\medskip
{\bf Event selection.}
If $\ln \Lambda_2$ is not negligible, not only is significance degraded but more importantly the expansion \Eq{eq:lnlami_exp} becomes unreliable; higher-order noise contributions cannot be reliably ignored then. Thus, we select only events with small $\ln \Lambda_2^i$
\beq
	\SNR_i>8, \quad \ln\Lambda^i_0 \geq \ln\Lambda^i_2.
\label{eq:selection}\eeq
The first condition is required for the detection of GW itself, let alone lensing effects. Although we assume here that GW signals can be detected as long as they satisfy the $\SNR > 8$ criterion, this assumption may not hold for some rare, strongly lensed signals (see Ref. \cite{Chan:2024qmb}). However, since our analysis primarily focuses on weakly lensed signals, this limitation has only a minor impact.

\medskip
{\bf Why variance?}
The total significance can be statistically related to the Variance of $\eta'$ and hence to the power spectrum
\begin{align}
	\left<\ln\Lambda \right> &\= \sum_i \left< \ln\Lambda^i \right> \,\simeq\, \sum_i \ln\Lambda^i_0 \,\sim\, \sum_i \SNR_i^2\, \left|\frac{d\eta_i}{d\ln f}\right|^2  \\
	&\,\simeq\, \overline{\rm SNR^2} \cdot \Var \left( \frac{d\eta}{d\ln f} \right) \,\propto\, \bar{k}_F^2 P(\bar{k}_F).
	\label{eq:lnlam_ave}
\end{align}
We assume that the Born approximation holds for the lensing effect. Hence, $\langle \eta' \rangle = 0$ and $\Var(\eta') = \langle |\eta'|^2 \rangle$.
The second line is valid if $N_F$ is large so that the CLT applies to $\eta'$; see \Sec{sec:mc} and \Fig{fig:var1}. This is of course intimately related to the Gaussianity of total $\ln \Lambda$; see \Sec{sec:properties}. If not Gaussian, again higher moments of $\eta'$ distribution matter so that the second line is not valid. 
In the first line we have used \Eq{eq:lnLambda0_approx}, and in the second line we have used that SNR$_i^2$ and $\eta'_i$ are independent with statistical mean $\overline{{\rm SNR}^2}$ and $0$, respectively. The last approximation in terms of $\bar{k}_F$ is the useful property discussed in \Sec{sec:multilensing}.

\Eq{eq:lnlam_ave} shows that the Variance of $\eta'$ can determine detection sensitivities. It is directly proportional to the $\ln \Lambda$ mean value. And the width of $\ln \Lambda$ Gaussian distribution becomes narrower as more events are added. Thus, the Variance alone becomes the most critical property determining $\ln \Lambda$ distributions, hence detection sensitivities.

\section{Monte-Carlo simulation} \label{sec:method}

\subsection{Event generation}  \label{sec:mc}

We use two Monte-Carlo event simulations for $\ln \Lambda_i$ distributions. One is based on a Gaussian probability distribution of gravitational potential $\tilde{\Phi}(\bk)$ (given by power $P(k)$), while the other on a random spatial distribution of point lenses. We validate our methods in the case where Gaussianity is expected to hold. We use following cosmological parameters: $h=0.6766$, $\bar{\rho}_0=1.27\times 10^{11} M_\odot/\Mpc^3$, $\Omega_m=0.3097$~\cite{Planck:2018vyg}.

\medskip
{\bf Method 1}.
The gravitational potential $\tilde{\Phi}(\bk)$ is assumed to follow a Gaussian probability distribution without higher-order moments, according to
\beq
\langle \, \tilde{\Phi}(\bk) \tilde{\Phi}^*(\bk') \,\rangle \= (2\pi)^3 P_{\Phi}(k) \,\delta^3(\bk-\bk'),
\eeq
where $P_\Phi(k) = \left(4\pi G \bar{\rho} a^{-1}/{k^2}\right)^2 P(k)$. For each event, random $\tilde{\Phi}(\bk)$ is generated, and $\eta_i(f)$ is calculated according to \Eq{eq:etaPk} (more precisely, \Eq{eq:etaPk_disc}).

For the sake of numerical simulation, we discretize $\bk_a$ with the lattice size $\Delta k$, and obtain a discrete set of values $\tilde{\Phi}(\bk_a)$ according to
\beq
\langle \, \tilde{\Phi}(\bk_a) \tilde{\Phi}^*(\bk_b) \,\rangle \= (2\pi)^3 P_\Phi(k) \,\frac{\delta_{ab}}{\Delta k^3}.
\label{eq:phi_sample}\eeq
Each $\tilde{\Phi}(\bk_a)$ is independent, even between the nearest neighbors. So the $1/\Delta k^3$ factor accounts for the statistical variance of stochastic functions in the discrete space. This is important because smooth functions can be approximated by the values at nearby discrete points, while stochastic functions cannot be. As a result, the integral formula for $\eta_i$ in \Eq{eq:etaPk} becomes a discrete summation over $\tilde{\Phi}(\bk_a)$ (sampled by \Eq{eq:phi_sample})
\begin{align}
	\eta(f; \chi_s) \,\simeq\, \sum_{a \,\in\, {\rm lattice}} \left( \frac{\Delta k}{2\pi} \right)^{3} \, \tilde{\Phi}(\bm{k}_a) \int_0^{\chi_s} d\chi\, e^{ik_{a,\parallel} \chi}  \, g(f; \bm{k}_a, \chi).
\label{eq:etaPk_disc}\end{align}

In general, however, matter distributions are not solely described by two-point correlations, but higher moments may well be non-vanishing.
Simulating such general mass distributions is one motivation for the second Method.

\medskip
{\bf Method 2}.
In this method, the potential field $\tilde{\Phi}(k)$ is simulated from a random spatial distribution of discrete lenses. This method is more general and correct than Method 1, but is computationally resource consuming. So we use this method only for PBH point-lens, in which the Gaussianity issue is most relevant.

The random spatial distribution of point lenses has two-point correlation given by the constant shot-noise power $P(k)$\footnote{We ignore possible PBH structures, such as isocurvature fluctuations and PBH halos.}
\beq
	P(k)_{\rm PBH} \= \frac{f_{\rm PBH}^2}{n_{\rm PBH}}.
\eeq
The power is not needed in this method, but can be used to compare with Method 1.
In general, random distributions have higher moments as well. But still, GW observables from Method 2 can agree with those from Method 1, as will be discussed thoroughly.

\begin{figure}[t]
	\centering
	\includegraphics[width=.45\linewidth]{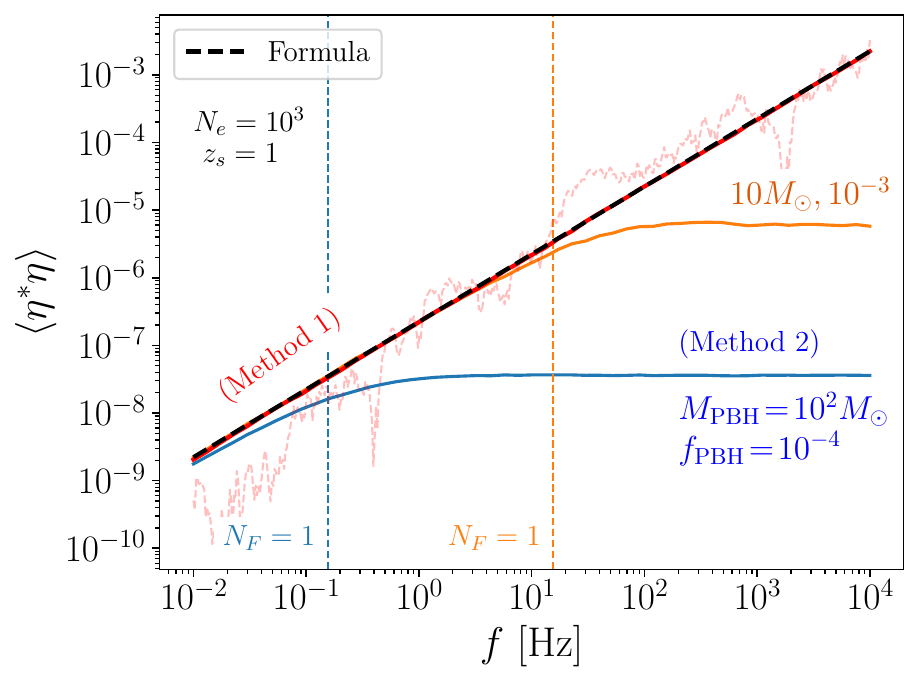}
    \includegraphics[width=.45\linewidth]{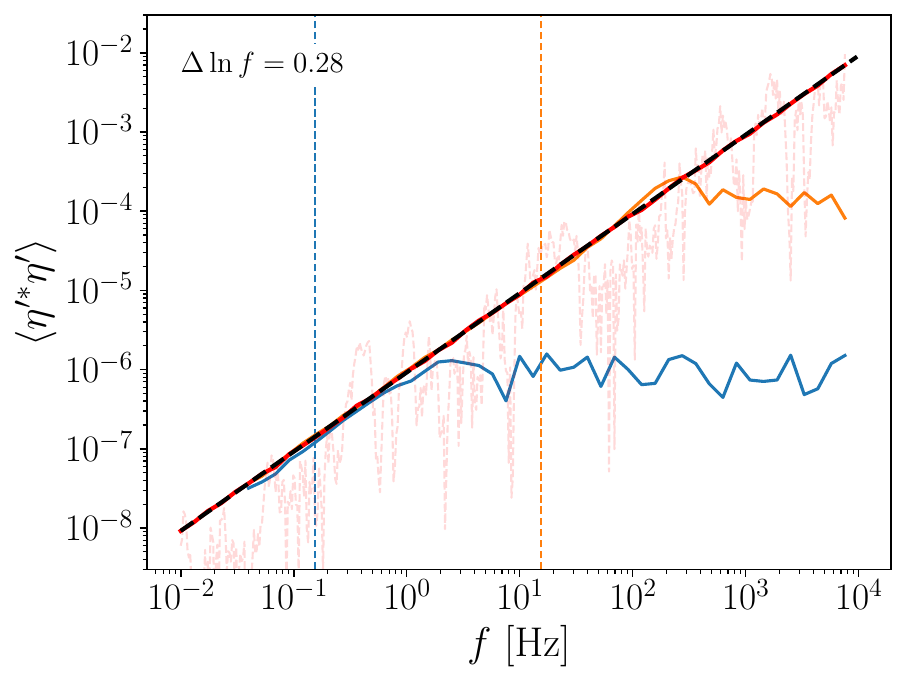}
	\caption{The Variance of $\eta$ (left) and $\eta'$ (right) simulated by Method 1 (red), Method 2 (blue, orange), and calculated by \Eq{eq:var_eta} (dashed). Method 1 always predicts Gaussian distributions, agreeing with \Eq{eq:var_eta}. Method 2 deviates from these in the high-frequency regime delineated by $N_F \lesssim 1$; two results of Method 2 having the same power and different $N_F$ deviate at respective frequencies. See \Sec{sec:mc} and \ref{sec:properties}. Also shown is one example realization by Method 1.}
	\label{fig:var1}
\end{figure}

\medskip
{\bf $\eta'$ ensembles from Method 1 versus 2.}
The statistical variance of GW observable $\eta'$, calculated from an ensemble of events generated by Method 1 and 2, are compared in \Fig{fig:var1}. The $\eta'$ distribution reflects that of $\tilde{\Phi}(\bk)$, from \Eq{eq:etaPk_disc}. Thus, the distribution is always Gaussian in Method 1, while not generally in Method 2. But the figure shows that the distribution from Method 2 also resembles the Gaussian case in the parameter space characterized by large $N_F \gtrsim 1$. This is because CLT applies to the ensemble of $\eta'$ generated by Method 2, when there are enough number of elements that affect $\eta'$ sizably.

The relevance of $N_F$ is further supported by the two cases of Method 2 shown in the figure. The two cases have the same power ($P \propto f_{\rm PBH} M_{\rm PBH}$) but different $M_{\rm PBH}$ and number densities. Thus, $N_F \propto f_{\rm PBH}/M_{\rm PBH}$ is different. Consequently, the two cases exhibit Gaussianity in different regimes, delineated by respective $N_F \sim 1$.

When the GW wavelength is shorter than the Schwarzschild radius (i.e., when the Einstein radius $r_E$ exceeds the Fresnel scale $r_F$), post-Born effects can become significant~\cite{Mizuno:2022xxp}.
These higher-order corrections may induce deviations between Method 1 and Method 2 even when $N_F \gtrsim 1$.
Such effects are suppressed if $N_F \lesssim 1$, where the non-Gaussian effects in Method 2 already dominate, making the post-Born corrections less significant.
We found that these effects become prominent only at very high frequencies and for low-mass PBHs, conditions that lie practically outside the scope of this study.
Therefore, we primarily discuss the lensing effect in the scope of the Born approximation in this study.
A detailed study incorporating post-Born corrections is left for future work.

We emphasize that the statistical results from Method 2 depend on the spatial variation of overdensities, not on the mass density itself. This means that our observable is determined by the variance of overdensities, as characterized by the power. For instance, increasing the PBH number density $n$, while keeping the total mass density $\rho$ constant eventually leads to zero power with a uniform distribution, since $P \propto \rho/n$. Instead, Method 2 approaches to Method 1 if the power is kept constant while $n$ increases (and $\rho$ varies too). Only cases with the same power can be reasonably compared in our study, as they are the ones that induce the same observables.

\subsection{GW waveforms, sources, and detectors}

\begin{figure}[t]
	\centering
	\includegraphics[width=.5\linewidth]{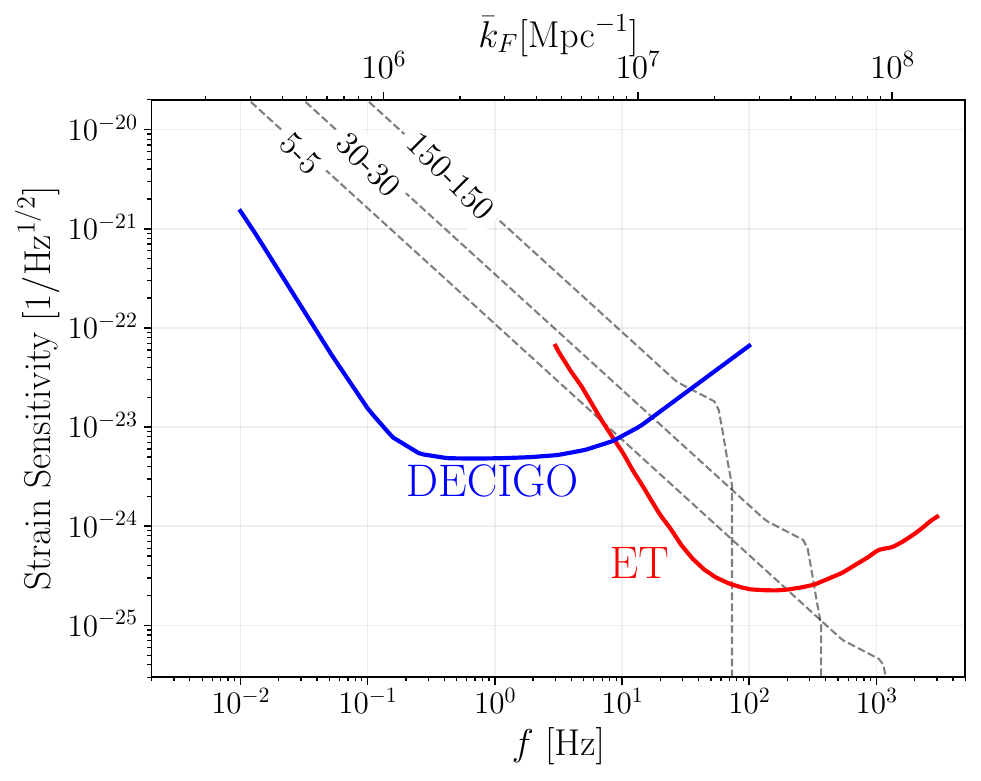}
	\caption{
		Sensitivity curves $S_n(f)$ for Einstein Telescope~\cite{Punturo:2010zz, Hild:2010id} and DECIGO~\cite{Kawamura:2020pcg}.
		Also shown are example waveforms from equal-mass binary mergers.
		The upper horizontal axis shows $\bar{k}_F(f)$  (with $z=1$) corresponding to $f$ on the lower horizontal.
	}
	\label{fig:Snf}
\end{figure}

For chirping waveforms, we include inspiral, merger, and ringdown phases. The merger and ringdown phases are also relevant albeit short, because they lead to non-negligible SNR as well as non-trivial frequency dependences.
We use \texttt{IMRPhenomA} waveform template~\cite{Ajith:2007kx}.
It is a phenomenological model for non-spinning binary mergers, which is simple and optimistic.

The waveform at the leading order is
\begin{align}
    h(f) \= A_p A(f) A_0 e^{i(2\pi f t_c^0 +\phi_c^0 + \Psi(f))},
    \label{eq:hf}
\end{align}
with $A(f)$ containing the chirping evolution in inspiral, merger, and ringdown phases
\begin{align}
	A(f) \= \left
		\{
		\begin{matrix}
			A_{\rm insp}(f) && f < f_{\rm merg} && {\rm (inspiral)} \\
			A(f_{\rm merg}) \, \left( \frac{f}{f_{\rm merg}} \right)^{-2/3}  && f_{\rm merg} < f < f_{\rm ring} && {\rm (merger)} \\
			A(f_{\rm ring}) \, \frac{\sigma_f^2/4}{(f-f_{\rm ring})^2+\sigma_f^2/4}   && f_{\rm ring} < f < f_{\rm cut} && {\rm (ringdown)}  
		\end{matrix}
	\right. ,
\end{align}
where
\begin{align}
    A_{\rm insp} (f) \= \sqrt{\frac{5}{96}} \frac{ {\cal M}^{5/6} f^{-7/6} \pi^{-2/3} }{D_s}. 
\end{align}
$D_s$ is the luminosity distance to the source.
$A_p$ factor includes polarization angle, binary inclination, and detector orientation dependences.
For simplicity, we set $A_p=1$.
The detailed expressions for $f_{\rm merg, \,ring, \, cut}$ and $\sigma_f$ are collected in \cite{Ajith:2007kx}.

For the population of binary black holes, we consider three parameters: $M_{\rm BBH}$, $\eta_{\rm BBH}$, and $z_{\rm BBH}$.
The distribution of the source frame total mass ($M_{\rm BBH}$) and the symmetric mass ratio ($\eta_{\rm BBH}$) are taken from \cite{Talbot:2018cva}.
Binary black holes are assumed to be uniformly distributed in the comoving coordinate up to the redshift $z_{\rm BBH}=10$.
The total merger rate is normalized to $R_0=28.3 \, \Gpc^{-3}{\rm yr}^{-1}$.

We use Einstein Telescope (ET) and DECIGO as benchmark missions.
The power spectral density $S_n(f)$ for each detector is shown in \Fig{fig:Snf}. The frequency $f=f_0$ from which SNR contribution is largest is $f_0\simeq 40$ and 0.15 Hz, respectively; we sometime use this $f_0$ for quick estimation.
Using $f_0$ in \Eq{eq:bkF}, one can expect that ET and DECIGO will be sensitive to structures at subgalactic scales $k = {\cal O}(0.01 - 10) \,{\rm pc}^{-1}$.

\begin{figure}[t]
	\centering
	\includegraphics[width=.5\linewidth]{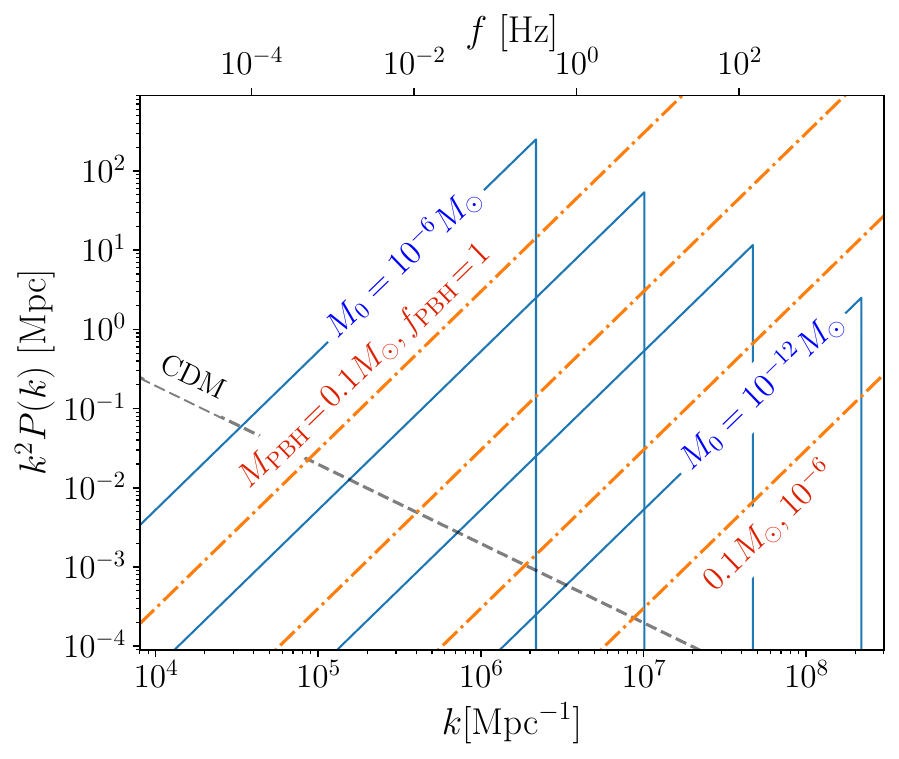}
	\caption{
		Model power spectra $k^2P(k)$ at $z=0$ for cold dark matter CDM (dashed), PBHs with $0.1M_\odot$ (dot-dashed), and axion minihalos from isocurvature perturbations (solid). See text for model parameters. 
		The frequency $f$ on the upper horizontal axis corresponds to $k$ on the lower horizontal by the relation $k=\bar{k}_F(f)$ (\Eq{eq:bkF}).
	}
	\label{fig:powerspec}
\end{figure}

\section{Sensitivities on models of small-scale structure} \label{sec:result}

We present final sensitivities on PBHs, axion minihalos, and general power spectrum. Model power spectra, $k^2P(k)$ relevant to the Variances, are shown in \Fig{fig:powerspec}. Along with PBH results, we also scrutinize statistical properties of multi-lensing: Gaussianity of $\ln \Lambda$, importance of sub-critical events, and importance of multiple lenses along the line of sight.

\subsection{PBH, and statistical properties of multi-lensing} \label{sec:properties}

\begin{figure}[t]
    \centering
    \includegraphics[width=.49\linewidth]{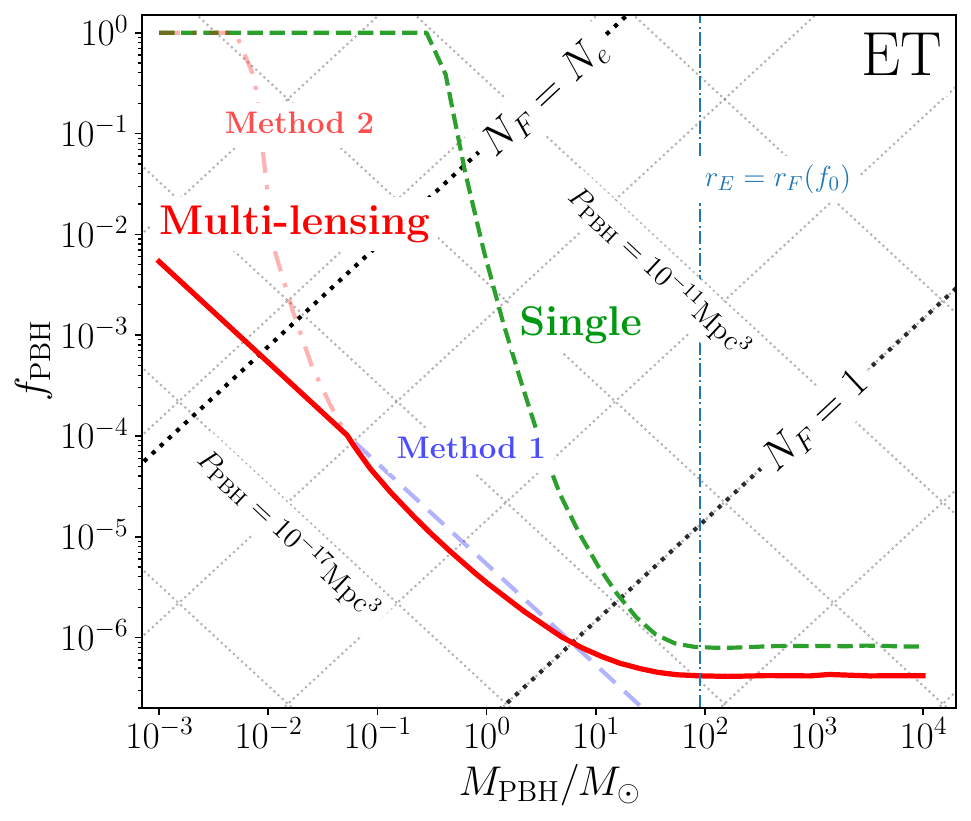}
    \includegraphics[width=.49\linewidth]{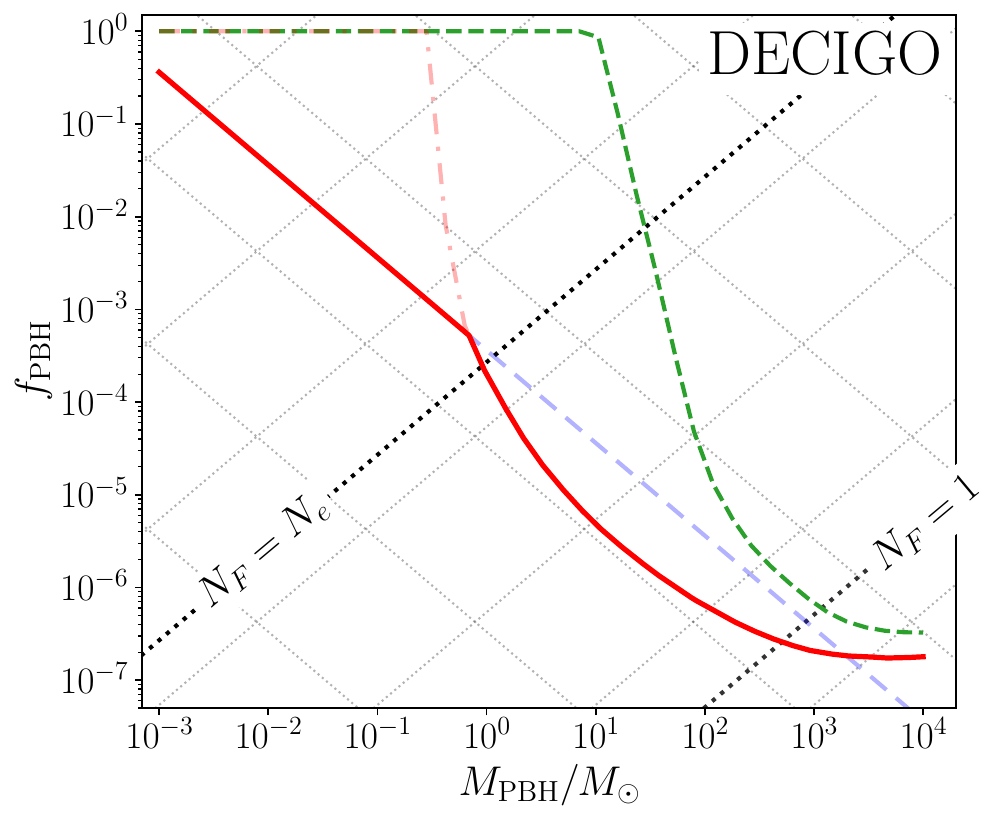}
    \includegraphics[width=.49\linewidth]{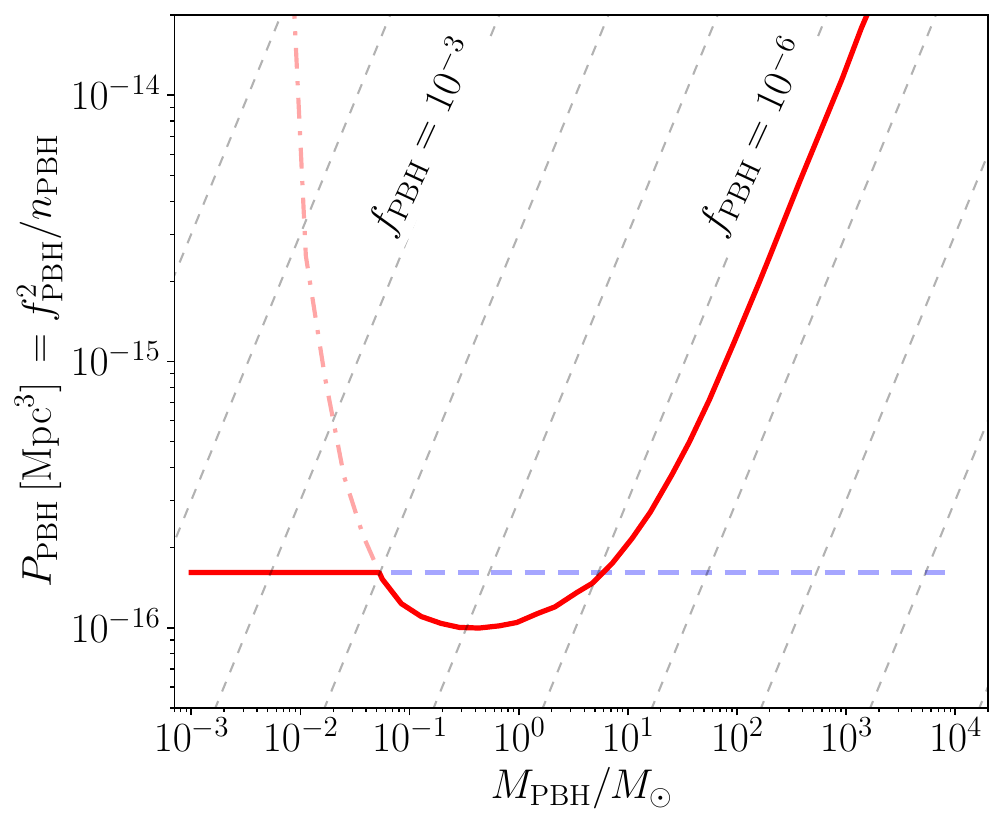}
    \includegraphics[width=.49\linewidth]{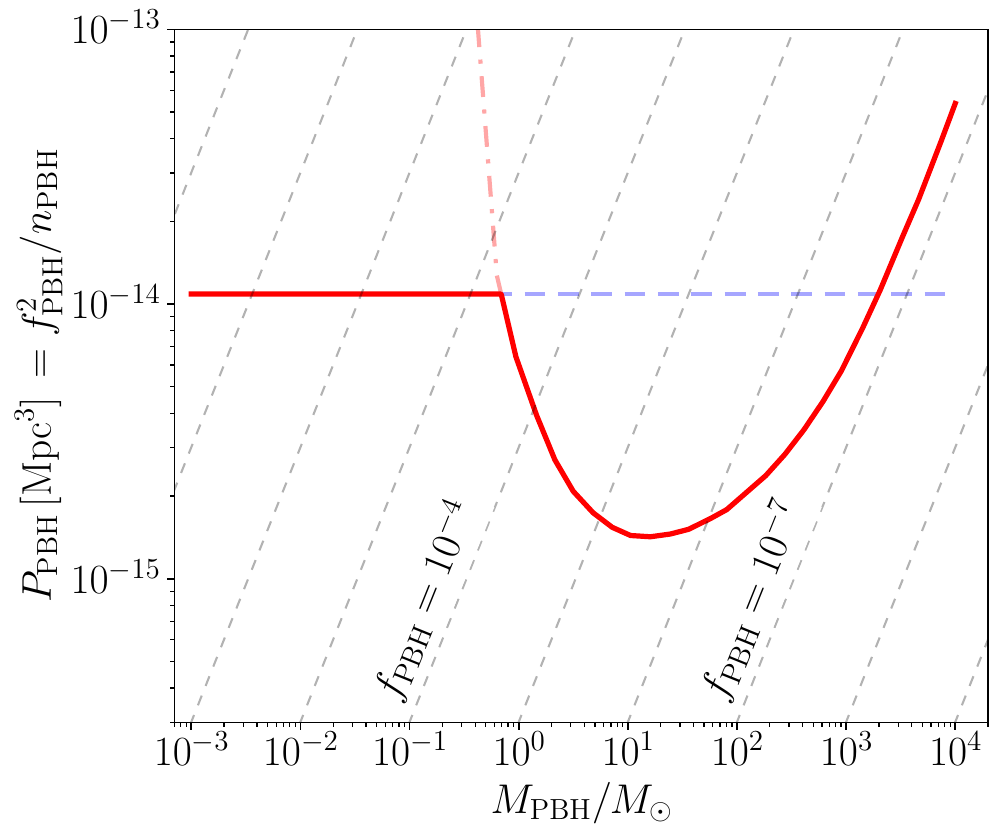}
    \caption{
        Sensitivities on PBHs from 5-year observations with ET (left) and DECIGO (right), in terms of the abundance $f_{\rm PBH}$  (upper) or constant power $P=f_{\rm PBH}^2/n_{\rm PBH}$ (lower). Our final results are red-solid lines, interpolated between Method 1 and 2; see text for detail. For comparison we also show projections of single strong-lensing event (green-dashed). Also shown are contours of $P_{\rm PBH}$, $N_F$, and the vertical mark at $r_E=r_F(f_0)$.
    }
    \label{fig:PBH_constraint_srcs}
\end{figure}

The presence of PBHs can contribute in several ways to the power spectrum, e.g. via their shot noise at short scale and their clustering affecting CDM halos at mid scale~\cite{Oguri:2020ldf,Afshordi:2003zb, Gong:2017sie, Inman:2019wvr}.
In this paper, we consider only the shot noise contribution as it dominates in the subgalactic scale $k=10^{5}\sim 10^{8}\Mpc^{-1}$ under our consideration.

As introduced in \Sec{sec:mc}, we simulate PBH lenses using two methods: Gaussian two-point correlation of potentials (Method 1) and random spatial distribution (Method 2). The two-point correlation for the random distribution is given by the constant shot noise
\beq
P_{\rm PBH} \= \frac{f_{\rm PBH}^2}{n_{\rm PBH}}.
\eeq

\medskip
{\bf Sensitivities on PBH.}
\Fig{fig:PBH_constraint_srcs} shows the final sensitivities on the PBH abundance $f_{\rm PBH}$ or on the constant power $P_{\rm PBH}$, from 5-year observations with ET or DECIGO. Our results with multi-lensing are shown as solid lines ($\ln \Lambda > 3$ with 90\% confidence), which is interpolated between the results of Method 1 and 2 as will be detailed. For comparison, we also show previous projections of single strong-lensing event (one event with $\ln \Lambda^i \geq 3$)~\cite{Jung:2017flg,GilChoi:2023ahp}\footnote{Our reproduction here is slightly stronger than the result in \cite{GilChoi:2023ahp} since we also consider merger and ringdown phases in the waveform, which contribute sizable SNR and frequency dependence.}. 

Our multi-lensing results are strong in the range $M_{\rm PBH} = 10^{-4} \sim 10^{2} \, M_\odot$ or in $k=10^6 \sim 10^8 \, {\rm Mpc}^{-1}$; the $k$-dependence will be clearer in other models discussed in next subsections. The results can significantly improve projections of single strong-lensing event~\cite{Jung:2017flg} consistently with \cite{Zumalacarregui:2024ocb} (LIGO-Virgo O3 have started to put constraints~\cite{LIGOScientific:2021izm,LIGOScientific:2023bwz}) and existing lensing constraints (currently constraining $f_{\rm PBH} \lesssim 10^{-1}$ for our mass range) from microlensing observations of nearby stars~\cite{EROS-2:2006ryy,CalchiNovati:2013jpj,DeRocco:2023hij}, caustic crossing of stars~\cite{Oguri:2017ock}, and type Ia supernovae lensing~\cite{Zumalacarregui:2017qqd}. The results are also competitive with other proposed statistical searches of fast-radio-burst lensing~\cite{Xiao:2024qay}.

Our results can be categorized into three regions, according to statistical properties of multi-lensing.
First, let us denote the total number of GW events during a given observation period by $N_e$.
In the middle region with $1 \lesssim N_F \lesssim N_e$, the result of Method 2 agrees with that of Method 1 so that the Gaussianity of $\ln \Lambda$ is achieved. As alluded, $N_F$ can be interpreted as counting the effective number of additions in $\ln\Lambda = \sum_i \ln \Lambda^i$, determining the validity of CLT for $\ln \Lambda$. In the rightmost region with $N_F \lesssim 1$, the Gaussianity is not expected, so Method 2 should be correct. Here, Method 2 rather asymptotes to the result of single strong event. In the leftmost region with $N_F \gtrsim N_e$, multiple lenses along the line of sight shall be relevant, on average. Here, Method 1 shall be correct as it effectively accounts for all lenses by potential simulation. Our final results interpolate these correct estimations in each region. We discuss each region in detail.

\begin{figure}[t]
	\centering
	\includegraphics[width=.5\linewidth]{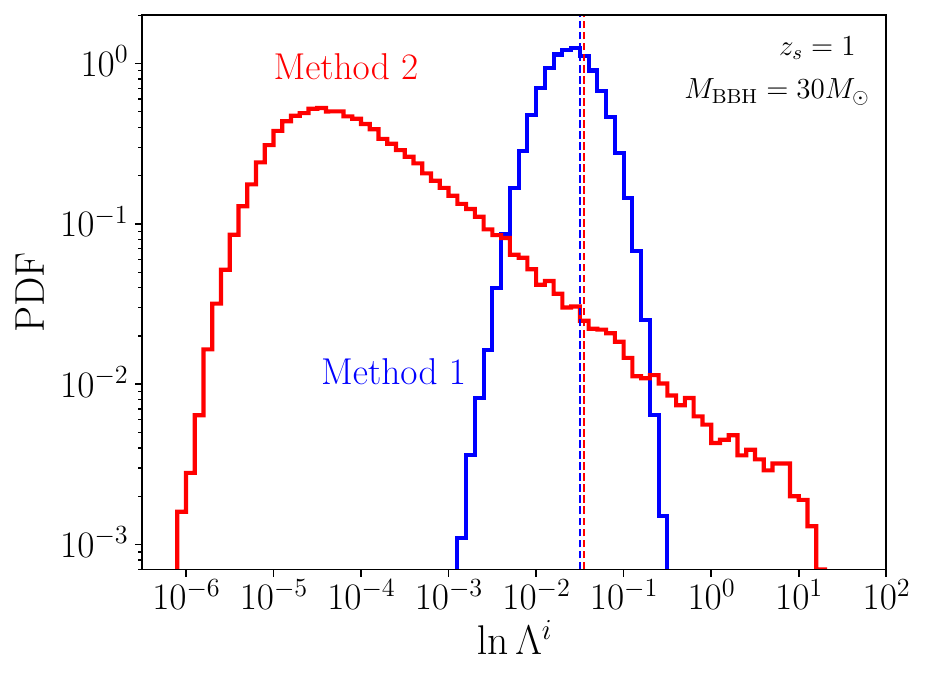}
	\caption{
		The probability distribution of individual $\ln \Lambda^i$ from Method 1 (blue) and Method 2 (red). It is well-localized distribution in Method 1. But in Method 2, it exhibits a long tail for large $\ln \Lambda^i$ when a point-lens happens to lie near the line of sight. The means of $\ln \Lambda^i$ (vertical lines) remain close between two methods. Source parameters are fixed for simplicity.
	}
	\label{fig:lnlami}
\end{figure}

\medskip
{\bf $N_F \lesssim 1$: non-Gaussian $\ln \Lambda$, and single strong event.} 
In the rightmost region with $N_F \lesssim 1$, the small $N_F$ means by definition that there is hardly a sizable diffraction. Most events exhibit very small $\ln\Lambda^i$, but in rare instances there can be strong lensing if a point lens happens to lie near the line of sight. Such a distribution of lensing strength $\ln \Lambda^i$ is highly non-local and non-Gaussian, as shown in \Fig{fig:lnlami}; although most events have small $\eta_i$, a long tail for large $\eta_i$ with small probability exists too. 

Thus, the sensitivity is determined dominantly by those rare events with strong lensing. This is exactly the previous proposals based on single strong-lensing event ($\ln \Lambda^i >3$ with 90\% confidence)~\cite{Jung:2017flg}. Our multi-lensing estimate (Method 2) approximately agrees with it in this region. But it also improves slightly due to some number of sub-critical events ($\ln \Lambda^i <3$ but sizable), which are not captured by single event searches. 

As the PBH mass increases, the Einstein radius grows, enhancing the effects of geometric optics~\cite{Choi:2021bkx,GilChoi:2023ahp}.
Geometrical optics effects become significant when the Einstein radius $r_E$ exceeds the Fresnel scale $r_F$.
However, for both ET and DECIGO, these effects are suppressed when $N_F \lesssim 1$ and can be neglected, as discussed in Sec.~\ref{sec:mc}.
Nevertheless, geometrical optics effects should be considered in other scenarios, such as those involving different detectors or lens objects.

\begin{figure}[t]
	\centering
	\includegraphics[width=.46\linewidth]{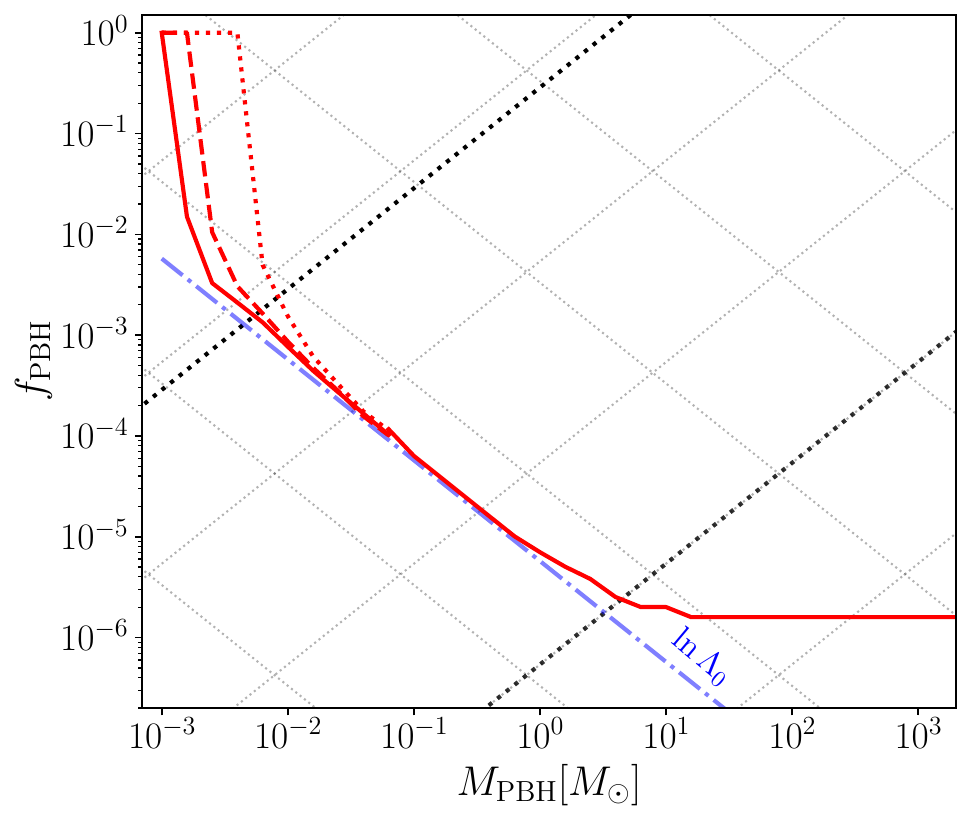}
	\includegraphics[width=.46\linewidth]{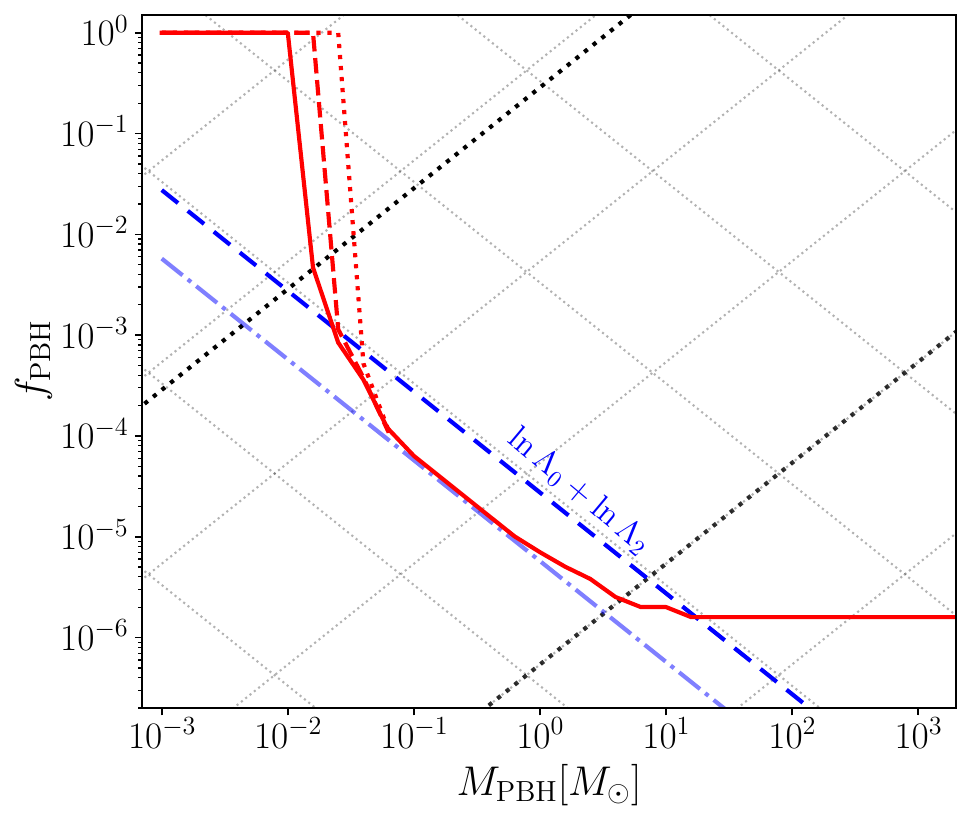}
	\caption{Sample sensitivity curves showing that in the region $N_F \gtrsim N_e$, multiple lenses are relevant on average (left), and $\ln \Lambda_2$ effects can be prominent on Method 2 (right). The results of summing first three lenses closest to the line of sight in Method 2 successively converge to Method 1. $\ln \Lambda_2$ affects Method 2 more sizably in $N_F \gtrsim N_e$, while Method 1 rather mildly and universally. $M_{\rm BBH}=15M_\odot$ and $z_s=1$ are fixed, for simplicity.
	}
	\label{fig:PBH_beta}
\end{figure}

\medskip
{\bf $N_F \gtrsim 1$: Gaussian $\ln \Lambda$, and many sub-critical events.}
In this region, the results of Method 1 and  2 agree, so the total $\ln \Lambda$ distribution is Gaussian even from Method 2. The sensitivity is determined by a sum of several sub-critical events with $\ln \Lambda^i < 3$, but it does not depend on any particular single event. An ensemble of $\ln \Lambda$ lacks extreme values and remains close to the mean value. Since the mean of $\ln \Lambda$ does not differ much between Method 1 and 2 (\Fig{fig:lnlami}), the sensitivities derived from both methods agree.

For $N_F \gtrsim 1$, the CLT guarantees the Gaussianity of $\ln \Lambda$, even though individual $\ln \Lambda^i$ is not (\Fig{fig:lnlami}). A related interpretation of $N_F$ was supported by \Fig{fig:var1}, in which the Gaussianity of $\eta'$ ensemble was achieved for $N_F \gtrsim 1$ even though individual $\eta'_i$ does not follow Gaussian distributions. Since $\ln \Lambda^i$ is approximately determined by $\eta'_i$, the same Gaussianity will be achieved similarly.

This region is where our multi-lensing can improve projections of single strong event by most. However, it is interesting to note that even with $N_F \gtrsim 1$ each event still experiences one or no lens within the Fresnel volume, on average.

\medskip
{\bf $N_F \gtrsim N_e$: Multiple lenses, and $\ln \Lambda_2$ effect.}
Genuine ``multi'' lensing events are, as shown in \Fig{fig:PBH_beta}, relevant to very large $N_F \gtrsim N_e$, in which on average more than one lenses are relevant in each event. In this region, Method 1 effectively accounts for all lenses by random potential simulation, while in Method 2 all relevant independent contributions of multiple lenses shall be added. \Fig{fig:PBH_beta} shows the convergence of Method 2 results toward the Method 1 as more number of lenses are added. 
Practically, it is challenging to add all discrete lenses in Method 2. Thus, we take the result of Method 1 as our final result in this region.

In this region, many number of sub-critical events sum up to yield sensitivities. Since each lensing signal is weak, $\ln \Lambda_2$ effect must be considered. As the right panel of \Fig{fig:PBH_beta} shows, the effect of $\ln \Lambda_2$ on Method 2 is most prominent in this region, degrading sensitivities. But  the effect on Method 1 is universal and rather mild in all three regions. So adding more discrete lenses in Method 2 is also critical to reduce $\ln \Lambda_2$ effect in this region, which is another reason to use Method 1 in this region.

\subsection{Sensitivities on axion minihalos}

We now obtain sensitivities on the axion in the scenario where isocurvature fluctuations lead to minihalos. Following standard axion cosmology, quantum fluctuations on the axion field during inflation become isocurvature density fluctuations after the axion field begins to oscillate as the Hubble time scale $1/H$ falls below the axion Compton time scale $1/m_a$. Once re-entering horizon, isocurvature fluctuations can evolve to form minihalos. Referring to \cite{Dai:2019lud,Fairbairn:2017sil} for details, we take the following simple model of minihalo power spectrum in this work, which can also be used with more realistic shape of power.

\begin{figure}[t]
	\centering
        \includegraphics[width=.6\linewidth]{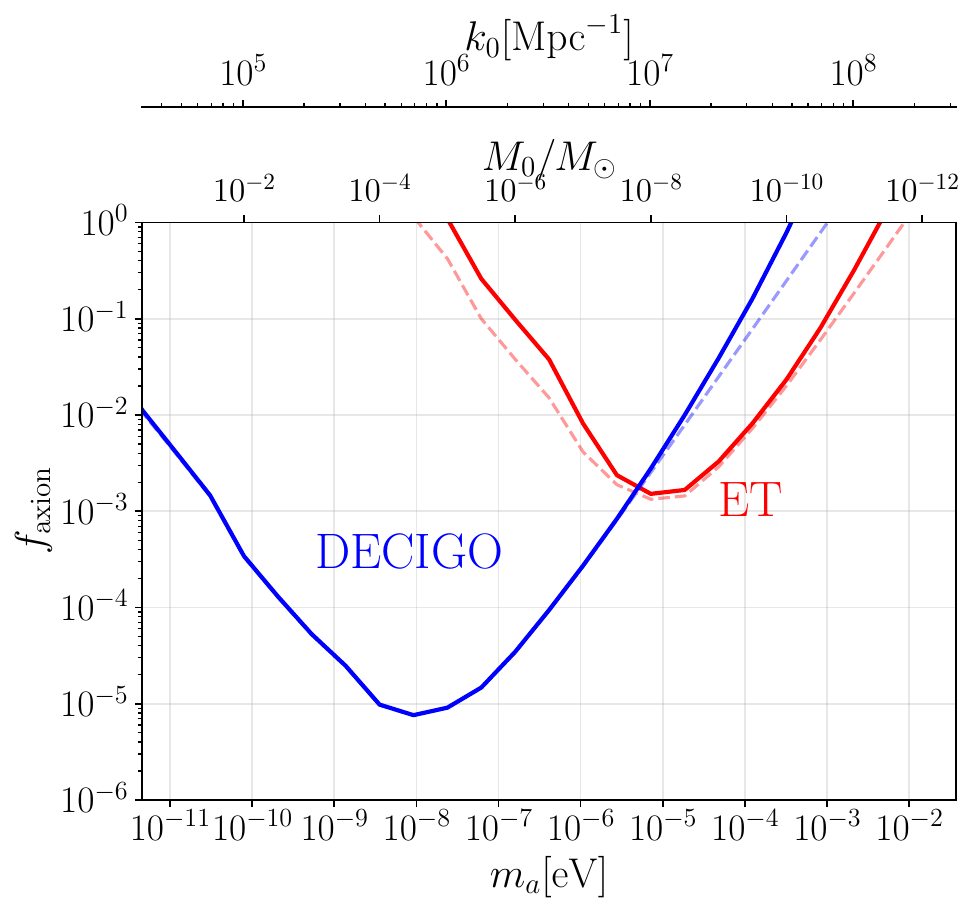}	
	\caption{
		The sensitivity on QCD axion minihalos from 5-year observation with ET(red) and DECIGO(blue), in terms of the axion abundance $f_{\rm axion}$. The results based on $\ln \Lambda_0 + \ln \Lambda_2$ (solid) as well as $\ln \Lambda_0$ (dashed) are shown. Method 1 is used since Gaussianity is expected.  On the upper horizontal axes, we also show corresponding minihalo mass $M_0$ and the peak scale $k_0$.
	}
	\label{fig:axion-result}
\end{figure}

In our simplified approach, the axion isocurvature power spectrum is approximated by a white-noise power spectrum \cite{Dai:2019lud, Hogan:1988mp, Fairbairn:2017sil}
\begin{align}
	P(k) = \Theta(k_0 - k) \frac{24\pi^2}{5k_0^3}\left( \frac{D_+(z)}{D_+(z_i)} \right)^2 \left( \frac{1+z_{\rm eq}}{1+z_i} \right)^2,
\end{align}
where the CDM growth factor $D_+(z) \simeq 1/(1+z)$ valid for $z\lesssim 10$ cancels the arbitrary $z_i$ dependence.
Here, $k_0 = a(t_0) H(t_0)$ is the cutoff on the scale, which is the comoving horizon at the time of axion oscillation $3H(t_0) \simeq m_a(t_0)$. For useful reference we use QCD axion mass with the temperature dependence taken from~\cite{Fairbairn:2017sil}. The total axion mass enclosed within the sphere with a radius $\pi/k_0$ sets the mass scale of minihalos
\begin{align}
	M_0 \= (4\pi/3) (\pi/k_0)^3 \bar{\rho}_{a0},  \qquad k_0 \,\simeq\, 7.55 \times 10^6 \,{\rm Mpc}^{-1} \, \left(\frac{10^{-8}\,{\rm M_\odot}}{M_0} \right)^{1/3}.
\label{eq:miniscale} \end{align}
where $\bar{\rho}_{a0}$ is the today's mean density of axions. We assume that all axions are in the form of minihalos with the mass $M_0$, and they constitute $f_{\rm axion}$ mass fraction of total dark matter.
Once realistic mass functions of minihalos and the fraction of axions in the form of minihalos are available, our baseline results can be corrected in a straightforward way.

The power spectrum is shown in \Fig{fig:powerspec}. The quantity that determines the statistical properties of lensing is $k^2 P(k)$. Notably, this exhibits a maximum at $k\simeq k_0$. Although realistic effects will smoothen the peak, the characteristic maximal nature around $k\simeq k_0$ will remain.

We simulate GW events using Method 1, assuming Gaussian potential fluctuations. This is a good approximation in this case (unlike some PBH cases). Since minihalos in the scale of interest are significantly lighter than the Solar mass, the mean Fresnel numbers of each event, $N_F/N_e \sim 10^9 \, (10^{-8}M_\odot/M_0) \, f_{\rm axion}$ for ET and $N_F/N_e \sim 10^{12} \,(10^{-8}M_\odot/M_0) \, f_{\rm axion}$ for DECIGO, are sufficiently large to assume Gaussian statistics. Thus, the power spectrum is all we need to obtain sensitivities. The effect of $\ln \Lambda_2$ will be mild as discussed.

\Fig{fig:axion-result} presents sensitivities on axion minihalos. ET is most sensitive to $M_0 \sim 10^{-8} \,M_\odot$ or $m_a \sim 10^{-5} \, {\rm eV}$, potentially down to $f_{\rm axion} \sim 10^{-3}$ with 5-year observations. These minihalos have $k_0 \sim 10^7 \,{\rm Mpc}^{-1}$ (\Eq{eq:miniscale}), to which ET's $f_0 \simeq 40$ Hz is most sensitive, according to the statistical properties of lensing (\Eq{eq:bkF}).
DECIGO, detecting about 100 times smaller frequencies, will be sensitive to 10 times larger scales (or 1000 times heavier minihalos).

\subsection{Sensitivities on general power spectrum}

We present some model-independent results that can be used to obtain sensitivities on particular models. Although GW observables in \Eq{eq:var_d_eta} are calculated by the convolution of power at various scales and distances, their statistical properties were dominated by a single scale $\bar{k}_F(f)$ for given $f$. This property could in principle be used to directly measure $P(k)$ from the observable at $f$; this opportunity is analogous to scanning the mass profile $\rho(r)$ of a single lens with the frequency spectrum of single GW diffraction event, based on the intuition portrayed in \Sec{sec:multilensing}. 

If possible, the result would have been really the model-independent \emph{measurement} of power spectrum function (or small scale structures). But in practice, the delta-function approximation by $k \sim \bar{k}_F(f)$ is not perfect, and a range of scales from a range of source distances is mixed in an ensemble. So the measurement at $f$ is not solely contributed by $P(\bar{k}_F(f))$ but by a mild convolution of $P(k)$ over some range of $k$ around $\bar{k}_F(f)$ (with some reference redshift $z=1$). 

In this section, we provide those numbers $\delta P(k)$ that can be used as a kernel of the convolution over $k$. Once a particular model of $P(k)$ is given, the sensitivity on the model can be obtained by the following convolution 
\beq
\ln \Lambda \,\simeq\, \ln \Lambda_c \, \int d \ln k\, \frac{P(k)}{\delta P(k)}.
\label{eq:delpk}\eeq
This is valid if $P(k)$ is not changing too rapidly with $k$. The Gaussianity or $N_F \gg 1$ also has to be assumed.

We obtain the numbers $\delta P(k)$ as follow. Assume a constant value $\delta P(k)$ of power only in a narrow bin of the size $\Delta k /k = 1$ around the given $k$. Find the value $\delta P(k)$ that yields $\ln \Lambda = \ln \Lambda_c = 3.0$. Method 1 has to be used, so that this result is valid for Gaussian cases. The effect of $\ln \Lambda_2$ will be mild as discussed.

\begin{figure}[t]
	\centering
	\includegraphics[width=0.6\linewidth]{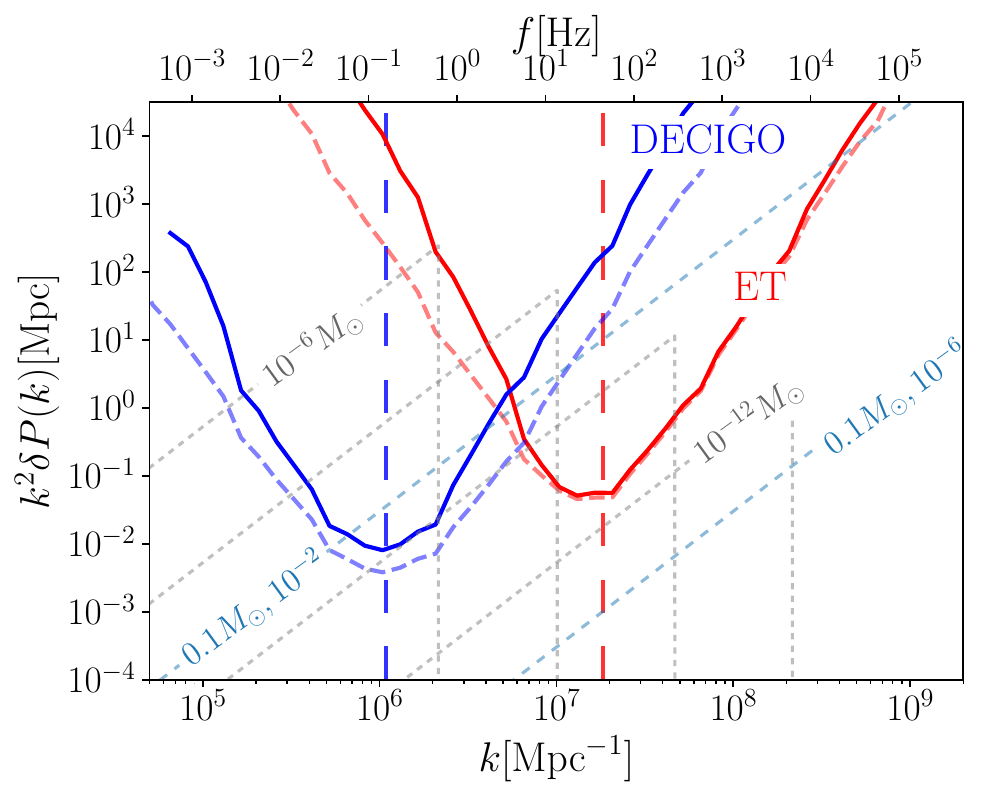}
	\caption{
		$\delta P(k)$ kernel for general sensitivities in \Eq{eq:delpk}, from 5-year observations with ET(red) and DECIGO(blue). One can convolute a given model $P(k)$ with $\delta P(k)$ to obtain the sensitivity. $\ln \Lambda_0+\ln \Lambda_2$ (solid) and $\ln \Lambda_0$ (dashed).
		For reference we also show some power spectra for PBHs and QCD axion minihalos.
		On the upper horizontal axis, we show the frequency by $k=\bar{k}_F(f)$ at $z=1$.
		$f=f_0$ (vertical).
	}
	\label{fig:Pk-sensitivity}
\end{figure}

\Fig{fig:Pk-sensitivity} presents these numbers $\delta P(k)$, from 5-year observations with ET and DECIGO. Again, this is not the final model-independent sensitivity, but only a kernel as in \Eq{eq:delpk}. However, this is still very useful since ${\cal G}_1$ is a narrow function, so that the sensitivity on $k$ is dominated by $\delta P(k)$.

Let us apply $\delta P(k)$ results to PBHs and axion minihalos, as examples. Near $k =k_{\rm ET} \simeq 10^{7} \,{\rm Mpc}^{-1}$ to which ET is most sensitive, $\delta P(k) \sim 10^{-16} \,{\rm Mpc}^{3}$.
With extra ${\cal O}(1)$ factors in the convolution, this roughly agrees with the constant PBH sensitivity shown in \Fig{fig:PBH_constraint_srcs} in the Gaussian regime $M_{\rm PBH} \lesssim 10^{-1} M_\odot$. 
Axion minihalos have power spectrum cut off at $k\lesssim k_0$, so only small enough halos with $k_0 \gtrsim k_{\rm ET}$ ($M_0 \lesssim 10^{-7} M_\odot$) will have non-negligible power probed by ET. This upper range of $M_0$ also roughly agrees with \Fig{fig:axion-result}, again up to ${\cal O}(1)$ factors. Similarly, the best sensitivity will be on $M_0 \sim 10^{-8} M_\odot$ having $k_0 \simeq k_{\rm ET}$ (recall that the minihalo $k^2 P(k)$ is maximal at $k \simeq k_0$), which also agrees with \Fig{fig:axion-result}.
Although the convolution between model $P(k)$ and $\delta P(k)$ has to be made for correct calculation, such an estimation based on peak values can be useful.

Our results on $\delta P(k)$ can be compared with another similar proposal based on the statistical variance of fast-radio-burst lensing~\cite{Xiao:2024qay}. Figure 4 therein shows the best sensitivity $k^2 \delta P(k) = 10^{-2} \sim 10^{-6} \,{\rm Mpc}$ at $k\sim 10^{8} \, {\rm Mpc}^{-1}$, and this is comparable to our best sensitivities $k^2 \delta P(k) = 10^{-1} \sim 10^{-2} \,{\rm Mpc}$ at slightly smaller $k = 10^{6}\sim 10^7 \,{\rm Mpc}^{-1}$. The scale of best sensitivity, in our study is determined by the relation $\bar{k}_F(f_0)$ in \Eq{eq:bkF}, while in that reference by year-long separated comparison $(v_{\rm DM} \cdot ({\rm yr}))^{-1} \sim 10^{8}\,{\rm Mpc}^{-1}$. Similarly, lower-frequency GWs can be tracked for year long duration to yield (1) a probe of larger scales (pc $\sim$ kpc), and (2) accumulated variances over the longer duration. We leave this for future study.

\section{Discussion}  \label{sec:summary}

We have worked out multi-lensing on chirping GWs by sub-galactic scale matter distributions. 
Through the relation \Eq{eq:bkF}, ET and DECIGO are expected to be sensitive to $k = 10^6 \sim 10^8$ and $10^5 \sim 10^7 \,{\rm Mpc}^{-1}$ scale, respectively, with the power down to 
$P(k) = 10^{-16}$ and $10^{-14} \,{\rm Mpc}^3$ level at the peak scale from 5-year observations. Applied to PBHs, which yield constant shot-noise $P(k)$ larger than CDM contributions in this range of $k$, the total abundance can be constrained by $f_{\rm PBH} \lesssim 10^{-6}$ for $M_{\rm PBH} = 1 M_\odot$. Applied to QCD axion minihalos, the total abundance can be constrained for $m_a = 10^{-7} \sim 10^{-3}$ and $10^{-12} \sim 10^{-4} \, {\rm eV}$ respectively, while assuming all axions are contained in minihalos. We have also obtained some model-independent results $\delta P(k)$, which can be combined with any given power $P(k)$ to yield the sensitivity on the model. 

The main property that allowed to understand the rather complicated 3d integral of lensing was the approximate property that statistical effects are dominated by a single scale $k \simeq \bar{k}_F(f_0) = k_F(f_0,\chi_l=\chi_s/2)$ \Eq{eq:bkF}. The physics underlying this property was discussed in \Fig{fig:G1}, which turned out to be consistent with single-lens intuitions that observable lensing effects are frequency dependencies induced by the shear while the total amplification depends on the total enclosed mass.
Having characteristic waveforms, GW was ideal to utilize these physics.

We have introduced the `Fresnel number' $N_F$ and shown that it is critical to statistical properties of multi-lensing. First of all, $N_F \sim 1, N_e$ delineated relative importances of sub-critical events and multiple lenses in an event. In addition, if the number density of dark structure is high, the total lensing likelihood is likely to be a sum of many effective lens contributions (albeit weakly lensed), so that the CLT guarantees that $\ln \Lambda$ follows a Gaussian distribution even though each $\ln \Lambda^i$ is far from Gaussian. In this case, the sensitivity obtained by random spatial distribution of lenses (Method 2) is same as the one obtained by Gaussian distribution of potentials (Method 1). Consequently, the sensitivity is essentially determined by the power spectrum among higher moments of overdensity distributions.

We have focused on relatively short final stage of inspiral and merger, with ET and DECIGO. By tracking  lower-frequency GWs for year-long duration, with LISA for example, one may be able to probe larger (but still sub-galactic) scales, by also utilizing statistical accumulation of lensing effects in a new way, i.e. accumulation of (in)dependent lensing effects on a single GW over longer periods. We leave this for future study.

Subgalactic scales remain as one of dark matter paradises, keeping its pristine properties. We expect future GW missions to shed light on this regime of the universe.

\acknowledgments
We thank Teruaki Suyama for his valuable comments.
SK and SJ are supported by Grant Korea NRF2019R1C1C1010050 and RS-2024-00342093. HGC is supported by the Institute for Basic Science (IBS) under the project code, IBS-R018-D3.

\newpage
\appendix
\section{Derivation of $\mathcal{G}_{0,1}$ kernels, dominated by $\chi_l \simeq \chi_s/2$} \label{app:G1}

\begin{figure}[h]
	\centering
	\includegraphics[width=.49\linewidth]{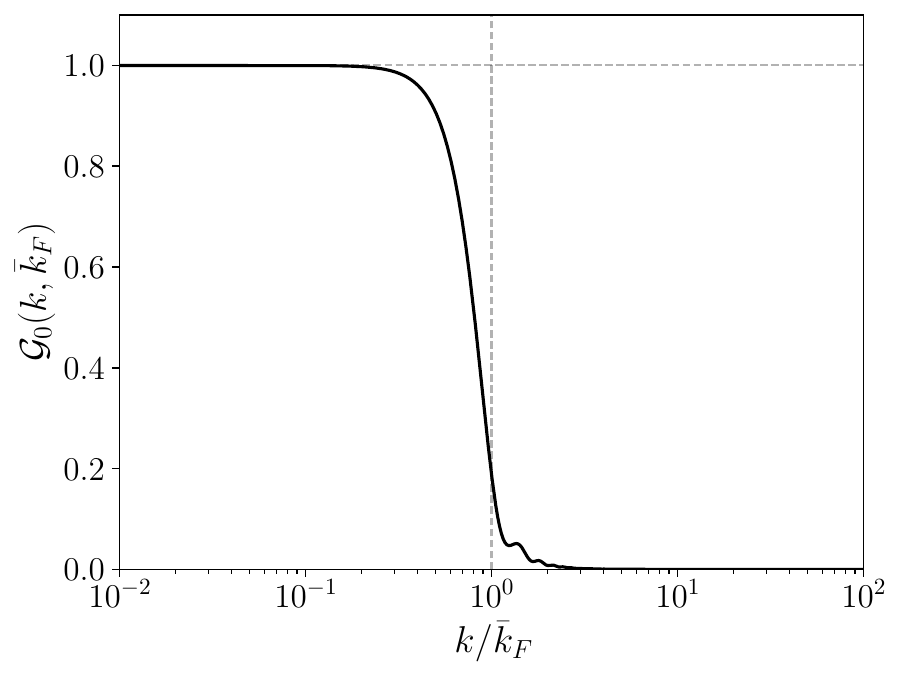}
	\caption{The $\mathcal{G}_0$ kernel for Var($\eta(f)$) at $z=0$ (\Eq{eq:G0_detail}). Since this Variance should measure the total enclosed mass density by the Fresnel radius, all and only modes $k \lesssim \bar{k}_F(f)$ contribute equally.
    }
	\label{fig:G0}
\end{figure}

In Eqs.~(\ref{eq:var_eta}) and (\ref{eq:var_d_eta}), we have defined the kernels ${\cal G}_{0,1}$ to absorb all $\chi_l$ dependent factors. In this appendix, we show that the resulting kernels are dominated by $\chi_l \simeq \chi_s/2$, which enables one of the main results of this paper: $k \simeq \bar{k}_F$ in Eqs.~(\ref{eq:G1delta}) and (\ref{eq:var_approx}).

The $\mathcal{G}_0$ kernel is rewritten and shown in \Fig{fig:G0}
\begin{align}
    \mathcal{G}_0(\ln k, \chi_s, f) &= \frac{960\pi^2 f^2}{k^4\chi_s^2}
	\int_0^{\chi_s} \frac{d\chi_l}{\chi_s} \, a(\chi_l)^{-2} \left[ 1 - \cos\left( \frac{\chi_l(\chi_s-\chi_l)}{4\pi f \chi_s} k^2 \right)\right] \\
    &=\frac{15}{\pi^4}\frac{\bar{k}_F^4}{k^4}\int_0^1 dx\, a(x\chi_s)^{-2} \left[1-\cos\left( 
        2\pi^2x(1-x)\frac{k^2}{\bar{k}_F^2}
    \right)\right],
    \label{eq:G0_detail}
\end{align}
where $x=\chi_l/\chi_s$ is introduced for simplicity. We assumed that $P(k)$ is independent on the redshift, but any dependence simply multiplies the $a(\chi_l)^{-2}$ factor and the same conclusion follows easily.

For $k\lesssim \bar{k}_F$, \Eq{eq:G0_detail} is approximated as
\begin{align}
    \mathcal{G}_0 &\,\simeq\, \frac{15}{\pi^4}\frac{\bar{k}_F^4}{k^4}\int_0^1 dx\, a^{-2} \left( 
        2\pi^2x(1-x)\frac{k^2}{\bar{k}_F^2}
    \right)^2 \\
    &\= \int_0^1 dx \, \frac{30x^2(1-x)^2}{a(x\chi_s)^2} \,\sim \,\frac{1}{a(\chi_s/2)^2}.
\end{align}
For $k \gtrsim \bar{k}_F$, only a saddle point of the cosine argument, $x=1/2$, contributes to the integral in \Eq{eq:G0_detail}. Thus, again, ${\cal G}_0 \sim a(\chi_s/2)^{-2}$. Again, any $\chi_l$ dependence of power simply modifies the final functional form, but not the dominance by $\chi_l \simeq \chi_s/2$.

As a result, up to the overall $\chi_l$-dependent factors, ${\cal G}_0$ approximately becomes a function only of $k$ and $\bar{k}_F$, which turns out to be a Heaviside function (consistent with Gauss' theorem as discussed). Therefore,
\begin{align}
    \mathcal{G}_0(k,\bar{k}_F) &\,\simeq\, a(\chi_s/2)^{-2} \, \Theta(\ln k_F - \ln k), \\
    \Var(\eta) &\,\propto\, \int_0^{\bar{k}_F} d\ln k\, k^2 P(k).
\end{align}
It directly indicates that the matter power spectrum contributes to the dispersion of amplifications with $k^2P(k)$ per $\ln k$ for $k<\bar{k}_F$.
In contrast, $P(k)$ in $k>\bar{k}_F$ does not contribute to the dispersion.

$\mathcal{G}_1$ kernel is more complicated because finite width 
 $\Delta \ln f$ is practically needed for careful consideration of derivative (see \Fig{fig:G1}). 
Similarly to \Eq{eq:var_d_eta} but including $\Delta \ln f$ carefully, 
\begin{align}
	{\rm Var} \left( \frac{\Delta\eta(f)}{\Delta \ln f} \right)
	&\,\simeq\, \frac{1}{2\pi} \int d\ln k \, k^2 P(k) \frac{1}{k^4} \int_0^{\chi_s} d\chi \, \left( \frac{4\pi G\bar{\rho}}{a} \right)^2 \left|\frac{\Delta g(f; k, \chi)}{\Delta \ln f}\right|^2  \nonumber \\
	&\,\equiv\, \frac{\chi_s^3}{60\pi}(4\pi G \bar{\rho})^2 \int d\ln k \, k^2 P(k) \, \mathcal{G}_1 (\ln k, \ln \bar{k}_F, \Delta \ln f) .
	\label{eq:var_d_eta2}
\end{align}
We assume $\Delta \ln f \lesssim 1$ here.
$g$ is defined as
\begin{align}
	g(f;k, \chi) &\= -i 4\pi f \left[ e^{-iu} - 1 \right], \\
	u &\= \frac{\chi(\chi_s-\chi)}{ 4 \pi  f \chi_s } k^2 \= \left(\frac{k}{\overline{k}_F}\right)^2 \frac{2\pi^2 \chi(\chi_s-\chi)}{\chi_s^2}.
\end{align}

First, consider $|u(f+\Delta f)-u(f)| = u\Delta \ln f \lesssim 1$ regime, where $\frac{\Delta g}{\Delta \ln f}\simeq \frac{dg}{d \ln f}$ holds well.
\begin{align}
	\left| \frac{dg}{d \ln f} \right|^2 &\= 32\pi^2 f^2 \left( 1-\cos u - u \sin u + \frac{u^2}{2}  \right), \\
	&\=\begin{cases}
            4\pi^2 f^2 u^4, & \text{if } u \ll \pi, \\
		16\pi^2 f^2 u^2, & \text{if } u \gg \pi.
	\end{cases}
	\label{eq:dg_dlnf}
\end{align}

Applying \Eq{eq:dg_dlnf} to \Eq{eq:var_d_eta2}, we obtain
\begin{align}
	\mathcal{G}_1 &\simeq  a(\chi_s/2)^{-2} \times
	\begin{cases}
		\frac{\pi^4}{21}\left(\frac{k}{\bar{k}_F}\right)^4, & \text{if } \bar{k}_F \gg k, \\
		1, & \text{if }  \bar{k}_F \ll k \ll \bar{k}_F/\Delta \ln f^{1/2}.
	\end{cases}
	\label{eq:G1_part1}
\end{align}

Now, consider $u\Delta \ln f \gtrsim 1$.
In this regime, $\frac{\Delta g}{\Delta \ln f}\simeq \frac{dg}{d \ln f}$ approximation does not hold anymore.
\begin{align}
	\left| \frac{\Delta g}{\Delta \ln f}\right|^2 &\,\simeq \,64\pi^2 f^2 \frac{\sin\left( \frac{1}{2}u \Delta \ln f \right)}{\Delta \ln f ^2}, \\
	\mathcal{G}_1 &\,\simeq\, \frac{15/\pi^{4}}{ a(\chi_s/2)^2 \Delta \ln f^{2}} \left(\frac{k}{\bar{k}_F}\right)^{-4}, \quad \text{if }  k \gg \overline{k}_F/\Delta \ln f^{1/2}.
	\label{eq:G1_part2}
\end{align}

From Eqs.~(\ref{eq:G1_part1}) and (\ref{eq:G1_part2}), by choosing proper $\Delta \ln f \sim {\cal O}(1)$ (as discussed in \Fig{fig:G1}) we can approximate $\mathcal{G}_1$ as a sharp function localized at $k\simeq \bar{k}_F$,
\begin{align}
	\mathcal{G}_1 &\,\sim\, \ln \left[\Delta \ln f^{-\frac{1}{2}}\right] \,\delta (\ln k-\ln \bar{k}_F).
\end{align}
This concludes one of the main results of this paper: statistical lensing effects are dominated by $k \simeq \bar{k}_F$.

\section{Formula for $\ln\Lambda$} \label{app:lnlam}

In this section, we present detailed formula for $\ln\Lambda_0$ and $\ln\Lambda_2$.
Consider a lensed GW wave data,
\begin{align}
    d \= h_L+n \= h_0(1+\eta) + n,
\end{align}
where $h_0$ is an unlensed wave and $F=1+\eta$ is a lensing amplification.
The detection likelihood function for the data can be obtained by~\cite{Choi:2021bkx}
\begin{align}
	\ln\Lambda \= \frac{1}{2} \left( \rho_{mL}^2 - \rho_{m0}^2 \right),
\end{align}
The matched filter SNRs for corresponding unlensed and lensed waveforms $h=h_0, h_L$ can be obtained by
\begin{align}
	\rho_{m}^2 &\= \max_{h} \left[ \left( d|d \right) - \left( d-h | d-h \right) \right] \\
		&\,\simeq\, \frac{1}{(h|h)} \left[ 
			(d|h)^2 + (d|ih)^2 +
			\frac{
				\left\{ (d|h)(d|i\omega h) - (d|ih)(d|\omega h) \right\}^2
			}{
				(d|h)(d|\omega^2 h) - (d|\omega h)^2 + (d|ih)(d|i\omega^2 h) - (d|i\omega h)^2
			}
		\right]
		\label{eq:rho_m}
\end{align}
where $\omega=2\pi f$.
The expression is derived by fitting the lensed and unlensed template waveforms $ h_{TL,0} = h_{L,0} \cdot A_c \exp[i\phi_c + i 2\pi f t_c] $ with three fitting parameters: $A_c$, $\phi_c$, and $t_c$, through the maximization in $\rho_{mL,0}^2$.
Each of three terms in Eq.~\ref{eq:rho_m} corresponds to the fitting of $A_c$, $\phi_c$, and $t_c$, respectively.
When $(n|h)/(h|h)\lesssim 1$, $\ln\Lambda$ can be expanded in combinations of $(n|\cdots h_{0,L})$.

In this study, it is sufficient to consider only the noise ensemble average of $\ln\Lambda$, instead of individual $\ln\Lambda$.
The noise average of the leading term becomes
\begin{align}
	\ln\Lambda_0 \= \left< \ln\Lambda \right>_{{\cal O}(n^0)} \= \frac{1}{2} \left( \rho_{mL}^2 - \rho_{m0}^2 \right)_{\substack{d=d_L}}
\end{align}

Since the linear terms become zero after taking average over the noise ensemble, one may ignore it.
\begin{align}
    \ln\Lambda_1 \= \left< \ln\Lambda \right>_{{\cal O}(n^1)} \= 0    
\end{align}
The sub-leading contribution comes from the second order term,
\begin{align}
	\ln\Lambda_2 \= \left< \ln\Lambda \right>_{{\cal O}(n^2)}.
\end{align}
By expanding \Eq{eq:rho_m} in $n$ and gathering the ${\cal O}(n^2)$ terms, $\ln\Lambda_2$ is obtained by the following formula for $h=h_{0,L}$,
\begin{align}
	&\ln \Lambda_2 \= \left< \mathcal{R}_L^2 \right> - \left< \mathcal{R}_0^2 \right>, \\
	&\left< \mathcal{R}^2 \right> \= 1 + \frac{1}{2(h|h)\mathcal{D}_0}(\nu_1 - \delta_1 \tau| \nu_1 - \delta_1 \tau), \\
	&\mathcal{D}_0 \= (h_L|h)(h_L|w^2h) - (h_L|wh)^2 + (h_L|ih)(h_L|iw^2h) - (h_L|iwh)^2, \\
	&\nu_1 \= (h_L|iwh)h - (h_L|wh)ih + (h_L|h)iwh - (h_L|ih)wh, \\
	&\delta^1 \= (h_L|w^2h)h + (h_L|iw^2h)ih + (h_L|ih)iw^2h + (h_L|h)w^2h - 2(h_L|iwh)iwh - 2(h_L|wh)wh, \\
	&\tau \= \frac{1}{\mathcal{D}_0} \left[ (h_L|h)(h_L|iwh) - (h_L|ih)(h_L|wh)  \right].
\end{align}
In our final results, we use these expressions for $\ln \Lambda_2$.


\begin{thebibliography}{99}


\bibitem{VanTilburg:2018ykj}
K.~Van Tilburg, A.~M.~Taki and N.~Weiner,
``Halometry from Astrometry,''
JCAP \textbf{07} (2018), 041
doi:10.1088/1475-7516/2018/07/041
[arXiv:1804.01991 [astro-ph.CO]].


\bibitem{Ando:2022tpj}
S.~Ando, N.~Hiroshima and K.~Ishiwata,
``Constraining the primordial curvature perturbation using dark matter substructure,''
Phys. Rev. D \textbf{106} (2022) no.10, 103014
doi:10.1103/PhysRevD.106.103014
[arXiv:2207.05747 [astro-ph.CO]].

\bibitem{Mondino:2023pnc}
C.~Mondino, A.~Tsantilas, A.~M.~Taki, K.~Van Tilburg and N.~Weiner,
``Astrometric weak lensing with Gaia DR3 and future catalogues: searches for dark matter substructure,''
Mon. Not. Roy. Astron. Soc. \textbf{531} (2024) no.1, 632-648
doi:10.1093/mnras/stae1017
[arXiv:2308.12330 [astro-ph.CO]].

\bibitem{Graham:2023unf}
P.~W.~Graham and H.~Ramani,
``Constraints on dark matter from dynamical heating of stars in ultrafaint dwarfs. I. MACHOs and primordial black holes,''
Phys. Rev. D \textbf{110} (2024) no.7, 075011
doi:10.1103/PhysRevD.110.075011
[arXiv:2311.07654 [hep-ph]].

\bibitem{Graham:2024hah}
P.~W.~Graham and H.~Ramani,
``Constraints on dark matter from dynamical heating of stars in ultrafaint dwarfs. II. Substructure and the primordial power spectrum,''
Phys. Rev. D \textbf{110} (2024) no.7, 075012
doi:10.1103/PhysRevD.110.075012
[arXiv:2404.01378 [hep-ph]].


\bibitem{Bai:2018bej}
Y.~Bai and N.~Orlofsky,
``Microlensing of X-ray Pulsars: a Method to Detect Primordial Black Hole Dark Matter,''
Phys. Rev. D \textbf{99} (2019) no.12, 123019
doi:10.1103/PhysRevD.99.123019
[arXiv:1812.01427 [astro-ph.HE]].

\bibitem{Dror:2019twh}
J.~A.~Dror, H.~Ramani, T.~Trickle and K.~M.~Zurek,
``Pulsar Timing Probes of Primordial Black Holes and Subhalos,''
Phys. Rev. D \textbf{100} (2019) no.2, 023003
doi:10.1103/PhysRevD.100.023003
[arXiv:1901.04490 [astro-ph.CO]].

\bibitem{Lee:2020wfn}
V.~S.~H.~Lee, A.~Mitridate, T.~Trickle and K.~M.~Zurek,
``Probing Small-Scale Power Spectra with Pulsar Timing Arrays,''
JHEP \textbf{06} (2021), 028
doi:10.1007/JHEP06(2021)028
[arXiv:2012.09857 [astro-ph.CO]].

\bibitem{Zumalacarregui:2017qqd}
M.~Zumalacarregui and U.~Seljak,
``Limits on stellar-mass compact objects as dark matter from gravitational lensing of type Ia supernovae,''
Phys. Rev. Lett. \textbf{121} (2018) no.14, 141101
doi:10.1103/PhysRevLett.121.141101
[arXiv:1712.02240 [astro-ph.CO]].


\bibitem{Munoz:2016tmg}
J.~B.~Mu\~noz, E.~D.~Kovetz, L.~Dai and M.~Kamionkowski,
``Lensing of Fast Radio Bursts as a Probe of Compact Dark Matter,''
Phys. Rev. Lett. \textbf{117} (2016) no.9, 091301
doi:10.1103/PhysRevLett.117.091301
[arXiv:1605.00008 [astro-ph.CO]].

\bibitem{Katz:2018zrn}
A.~Katz, J.~Kopp, S.~Sibiryakov and W.~Xue,
JCAP \textbf{12} (2018), 005
doi:10.1088/1475-7516/2018/12/005
[arXiv:1807.11495 [astro-ph.CO]].

\bibitem{Jung:2019fcs}
S.~Jung and T.~Kim,
``Gamma-ray burst lensing parallax: Closing the primordial black hole dark matter mass window,''
Phys. Rev. Res. \textbf{2} (2020) no.1, 013113
doi:10.1103/PhysRevResearch.2.013113
[arXiv:1908.00078 [astro-ph.CO]].

\bibitem{Xiao:2024qay}
H.~Xiao, L.~Dai and M.~McQuinn,
``Detecting dark matter substructures on small scales with fast radio bursts,''
Phys. Rev. D \textbf{110} (2024) no.2, 023516
doi:10.1103/PhysRevD.110.023516
[arXiv:2401.08862 [astro-ph.CO]].

\bibitem{Gould:1992}
A.~Gould,
``Femtolensing of gamma-ray bursters,''
  Astrophys.\ J.\  {\bf 386}, L5 (1992)
  doi:10.1086/186279

\bibitem{Nemiroff:1995ak}
R.~J.~Nemiroff and A.~Gould,
``Probing for MACHOs of mass $10^{-15}$-solar-mass - $10^{-7}$-solar-mass with gamma-ray burst parallax spacecraft,''
Astrophys. J. Lett. \textbf{452} (1995), L111
doi:10.1086/309722
[arXiv:astro-ph/9505019 [astro-ph]].

\bibitem{Nemiroff:2001bp} 
  R.~J.~Nemiroff, G.~F.~Marani, J.~P.~Norris and J.~T.~Bonnell,
  ``Limits on the cosmological abundance of supermassive compact objects from a millilensing search in gamma-ray burst data,''
  Phys.\ Rev.\ Lett.\  {\bf 86}, 580 (2001)
  doi:10.1103/PhysRevLett.86.580
  [astro-ph/0101488].


\bibitem{Niikura:2017zjd} 
  H.~Niikura {\it et al.},
  ``Microlensing constraints on primordial black holes with Subaru/HSC Andromeda observations,''
  Nat.\ Astron.\  {\bf 3}, no. 6, 524 (2019)
  doi:10.1038/s41550-019-0723-1
  [arXiv:1701.02151 [astro-ph.CO]].

\bibitem{Niikura:2019kqi} 
  H.~Niikura, M.~Takada, S.~Yokoyama, T.~Sumi and S.~Masaki,
  ``Constraints on Earth-mass primordial black holes from OGLE 5-year microlensing events,''
  Phys.\ Rev.\ D {\bf 99}, no. 8, 083503 (2019)
  doi:10.1103/PhysRevD.99.083503
  [arXiv:1901.07120 [astro-ph.CO]].

\bibitem{DeRocco:2023hij}
W.~DeRocco, N.~Smyth and V.~Takhistov,
``New Light on Dark Extended Lenses with the Roman Space Telescope,''
Astrophys. J. Lett. \textbf{965} (2024) no.1, L3
doi:10.3847/2041-8213/ad3644
[arXiv:2312.14782 [astro-ph.CO]].

\bibitem{EROS-2:2006ryy}
P.~Tisserand \textit{et al.} [EROS-2],
``Limits on the Macho Content of the Galactic Halo from the EROS-2 Survey of the Magellanic Clouds,''
Astron. Astrophys. \textbf{469} (2007), 387-404
doi:10.1051/0004-6361:20066017
[arXiv:astro-ph/0607207 [astro-ph]].

\bibitem{CalchiNovati:2013jpj}
S.~Calchi Novati, S.~Mirzoyan, P.~Jetzer and G.~Scarpetta,
``Microlensing towards the SMC: a new analysis of OGLE and EROS results,''
Mon. Not. Roy. Astron. Soc. \textbf{435} (2013), 1582
doi:10.1093/mnras/stt1402
[arXiv:1308.4281 [astro-ph.GA]].

\bibitem{Griest:2013aaa} 
  K.~Griest, A.~M.~Cieplak and M.~J.~Lehner,
  ``Experimental Limits on Primordial Black Hole Dark Matter from the First 2 yr of Kepler Data,''
  Astrophys.\ J.\  {\bf 786}, no. 2, 158 (2014)
  doi:10.1088/0004-637X/786/2/158
  [arXiv:1307.5798 [astro-ph.CO]].


\bibitem{Oguri:2017ock}
M.~Oguri, J.~M.~Diego, N.~Kaiser, P.~L.~Kelly and T.~Broadhurst,
``Understanding caustic crossings in giant arcs: characteristic scales, event rates, and constraints on compact dark matter,''
Phys. Rev. D \textbf{97} (2018) no.2, 023518
doi:10.1103/PhysRevD.97.023518
[arXiv:1710.00148 [astro-ph.CO]].

\bibitem{Dai:2019lud}
L.~Dai and J.~Miralda-Escud\'e,
``Gravitational Lensing Signatures of Axion Dark Matter Minihalos in Highly Magnified Stars,''
Astron. J. \textbf{159}, no.2, 49 (2020)
doi:10.3847/1538-3881/ab5e83
[arXiv:1908.01773 [astro-ph.CO]].



\bibitem{Jung:2017flg}
S.~Jung and C.~S.~Shin,
``Gravitational-Wave Fringes at LIGO: Detecting Compact Dark Matter by Gravitational Lensing,''
Phys. Rev. Lett. \textbf{122} (2019) no.4, 041103
doi:10.1103/PhysRevLett.122.041103
[arXiv:1712.01396 [astro-ph.CO]].

\bibitem{Dai:2018enj}
L.~Dai, S.~S.~Li, B.~Zackay, S.~Mao and Y.~Lu,
``Detecting Lensing-Induced Diffraction in Astrophysical Gravitational Waves,''
Phys. Rev. D \textbf{98} (2018) no.10, 104029
doi:10.1103/PhysRevD.98.104029
[arXiv:1810.00003 [gr-qc]].

\bibitem{Oguri:2020ldf}
M.~Oguri and R.~Takahashi,
``Probing Dark Low-mass Halos and Primordial Black Holes with Frequency-dependent Gravitational Lensing Dispersions of Gravitational Waves,''
Astrophys. J. \textbf{901}, no.1, 58 (2020)
doi:10.3847/1538-4357/abafab
[arXiv:2007.01936 [astro-ph.CO]].


\bibitem{GilChoi:2023ahp}
H.~Gil Choi, S.~Jung, P.~Lu and V.~Takhistov,
``Coexistence Test of Primordial Black Holes and Particle Dark Matter from Diffractive Lensing,''
Phys. Rev. Lett. \textbf{133} (2024) no.10, 101002
doi:10.1103/PhysRevLett.133.101002
[arXiv:2311.17829 [astro-ph.CO]].

\bibitem{Zumalacarregui:2024ocb}
M.~Zumalac\'arregui,
``Lens Stochastic Diffraction: A Signature of Compact Structures in Gravitational-Wave Data,''
[arXiv:2404.17405 [gr-qc]].

\bibitem{LIGOScientific:2021izm}
R.~Abbott \textit{et al.} [LIGO Scientific and VIRGO],
``Search for Lensing Signatures in the Gravitational-Wave Observations from the First Half of LIGO\textendash{}Virgo\textquoteright{}s Third Observing Run,''
Astrophys. J. \textbf{923} (2021) no.1, 14
doi:10.3847/1538-4357/ac23db
[arXiv:2105.06384 [gr-qc]].

\bibitem{LIGOScientific:2023bwz}
R.~Abbott \textit{et al.} [LIGO Scientific, KAGRA and VIRGO],
``Search for Gravitational-lensing Signatures in the Full Third Observing Run of the LIGO\textendash{}Virgo Network,''
Astrophys. J. \textbf{970} (2024) no.2, 191
doi:10.3847/1538-4357/ad3e83
[arXiv:2304.08393 [gr-qc]].

\bibitem{Nakamura:1997sw} 
  T.~T.~Nakamura,
  ``Gravitational lensing of gravitational waves from inspiraling binaries by a point mass lens,''
  Phys.\ Rev.\ Lett.\  {\bf 80}, 1138 (1998).
  doi:10.1103/PhysRevLett.80.1138

\bibitem{Chluba:2012we}
J.~Chluba, A.~L.~Erickcek and I.~Ben-Dayan,
``Probing the inflaton: Small-scale power spectrum constraints from measurements of the CMB energy spectrum,''
Astrophys. J. \textbf{758} (2012), 76
doi:10.1088/0004-637X/758/2/76
[arXiv:1203.2681 [astro-ph.CO]].

\bibitem{Cyr-Racine:2018htu}
F.~Y.~Cyr-Racine, C.~R.~Keeton and L.~A.~Moustakas,
``Beyond subhalos: Probing the collective effect of the Universe's small-scale structure with gravitational lensing,''
Phys. Rev. D \textbf{100} (2019) no.2, 023013
doi:10.1103/PhysRevD.100.023013
[arXiv:1806.07897 [astro-ph.CO]].


\bibitem{Choi:2021bkx}
H.~G.~Choi, C.~Park and S.~Jung,
``Small-scale shear: Peeling off diffuse subhalos with gravitational waves,''
Phys. Rev. D \textbf{104}, no.6, 063001 (2021)
doi:10.1103/PhysRevD.104.063001
[arXiv:2103.08618 [astro-ph.CO]].


\bibitem{Urrutia:2024pos}
J.~Urrutia and V.~Vaskonen,
``The dark timbre of gravitational waves,''
[arXiv:2402.16849 [gr-qc]].


\bibitem{Takahashi:2005sxa}
R.~Takahashi, T.~Suyama and S.~Michikoshi,
``Scattering of gravitational waves by the weak gravitational fields of lens objects,''
Astron. Astrophys. \textbf{438}, L5 (2005)
doi:10.1051/0004-6361:200500140
[arXiv:astro-ph/0503343 [astro-ph]].

\bibitem{Takahashi:2005ug}
R.~Takahashi,
``Amplitude and phase fluctuations for gravitational waves propagating through inhomogeneous mass distribution in the universe,''
Astrophys. J. \textbf{644}, 80-85 (2006)
doi:10.1086/503323
[arXiv:astro-ph/0511517 [astro-ph]].

\bibitem{Jung:2022tzn}
S.~Jung and S.~Kim,
``Solar diffraction of LIGO-band gravitational waves,''
JCAP \textbf{07}, 042 (2023)
doi:10.1088/1475-7516/2023/07/042
[arXiv:2210.02649 [astro-ph.CO]].


\bibitem{Chan:2024qmb}
J.~C.~L.~Chan, E.~Seo, A.~K.~Y.~Li, H.~Fong and J.~M.~Ezquiaga,
``Detectability of Lensed Gravitational Waves in Matched-Filtering Searches,''
[arXiv:2411.13058 [gr-qc]].


\bibitem{Planck:2018vyg}
N.~Aghanim \textit{et al.} [Planck],
``Planck 2018 results. VI. Cosmological parameters,''
Astron. Astrophys. \textbf{641}, A6 (2020)
[erratum: Astron. Astrophys. \textbf{652}, C4 (2021)]
doi:10.1051/0004-6361/201833910
[arXiv:1807.06209 [astro-ph.CO]].


\bibitem{Mizuno:2022xxp}
M.~Mizuno and T.~Suyama,
``Weak lensing of gravitational waves in wave optics: Beyond the Born approximation,''
Phys. Rev. D \textbf{108}, no.4, 043511 (2023)
doi:10.1103/PhysRevD.108.043511
[arXiv:2210.02062 [astro-ph.CO]].


\bibitem{Punturo:2010zz}
M.~Punturo, M.~Abernathy, F.~Acernese, B.~Allen, N.~Andersson, K.~Arun, F.~Barone, B.~Barr, M.~Barsuglia and M.~Beker, \textit{et al.}
``The Einstein Telescope: A third-generation gravitational wave observatory,''
Class. Quant. Grav. \textbf{27}, 194002 (2010)
doi:10.1088/0264-9381/27/19/194002

\bibitem{Hild:2010id}
S.~Hild, M.~Abernathy, F.~Acernese, P.~Amaro-Seoane, N.~Andersson, K.~Arun, F.~Barone, B.~Barr, M.~Barsuglia and M.~Beker, \textit{et al.}
``Sensitivity Studies for Third-Generation Gravitational Wave Observatories,''
Class. Quant. Grav. \textbf{28}, 094013 (2011)
doi:10.1088/0264-9381/28/9/094013
[arXiv:1012.0908 [gr-qc]].

\bibitem{Kawamura:2020pcg}
S.~Kawamura, M.~Ando, N.~Seto, S.~Sato, M.~Musha, I.~Kawano, J.~Yokoyama, T.~Tanaka, K.~Ioka and T.~Akutsu, \textit{et al.}
``Current status of space gravitational wave antenna DECIGO and B-DECIGO,''
PTEP \textbf{2021}, no.5, 05A105 (2021)
doi:10.1093/ptep/ptab019
[arXiv:2006.13545 [gr-qc]].



\bibitem{Ajith:2007kx}
P.~Ajith, S.~Babak, Y.~Chen, M.~Hewitson, B.~Krishnan, A.~M.~Sintes, J.~T.~Whelan, B.~Bruegmann, P.~Diener and N.~Dorband, \textit{et al.}
``A Template bank for gravitational waveforms from coalescing binary black holes. I. Non-spinning binaries,''
Phys. Rev. D \textbf{77}, 104017 (2008)
[erratum: Phys. Rev. D \textbf{79}, 129901 (2009)]
doi:10.1103/PhysRevD.77.104017
[arXiv:0710.2335 [gr-qc]].

\bibitem{Talbot:2018cva}
C.~Talbot and E.~Thrane,
``Measuring the binary black hole mass spectrum with an astrophysically motivated parameterization,''
Astrophys. J. \textbf{856}, no.2, 173 (2018)
doi:10.3847/1538-4357/aab34c
[arXiv:1801.02699 [astro-ph.HE]].



\bibitem{Afshordi:2003zb}
N.~Afshordi, P.~McDonald and D.~N.~Spergel,
``Primordial black holes as dark matter: The Power spectrum and evaporation of early structures,''
Astrophys. J. Lett. \textbf{594}, L71-L74 (2003)
doi:10.1086/378763
[arXiv:astro-ph/0302035 [astro-ph]].

\bibitem{Gong:2017sie}
J.~O.~Gong and N.~Kitajima,
``Small-scale structure and 21cm fluctuations by primordial black holes,''
JCAP \textbf{08}, 017 (2017)
doi:10.1088/1475-7516/2017/08/017
[arXiv:1704.04132 [astro-ph.CO]].

\bibitem{Inman:2019wvr}
D.~Inman and Y.~Ali-Ha\"\i{}moud,
``Early structure formation in primordial black hole cosmologies,''
Phys. Rev. D \textbf{100}, no.8, 083528 (2019)
doi:10.1103/PhysRevD.100.083528
[arXiv:1907.08129 [astro-ph.CO]].

\bibitem{Hogan:1988mp}
C.~J.~Hogan and M.~J.~Rees,
``AXION MINICLUSTERS,''
Phys. Lett. B \textbf{205}, 228-230 (1988)
doi:10.1016/0370-2693(88)91655-3

\bibitem{Fairbairn:2017sil}
M.~Fairbairn, D.~J.~E.~Marsh, J.~Quevillon and S.~Rozier,
``Structure formation and microlensing with axion miniclusters,''
Phys. Rev. D \textbf{97} (2018) no.8, 083502
doi:10.1103/PhysRevD.97.083502
[arXiv:1707.03310 [astro-ph.CO]].



\end{thebibliography}
\end{document}